 \documentclass[times,twocolumn,final,longtitle,monochrome]{elsarticle}

\usepackage{jasr}
\usepackage{framed,multirow}

\usepackage{amssymb}
\usepackage{latexsym}
\usepackage{siunitx}
\usepackage[gen]{eurosym}
\usepackage{float}
\usepackage{subcaption}

\usepackage{url}
\usepackage[table]{xcolor} 
\definecolor{newcolor}{rgb}{.8,.349,.1}

\usepackage[hidelinks]{hyperref}

\usepackage{cleveref}
\usepackage[para]{threeparttable}
\usepackage{etoolbox}

\AtEndEnvironment{threeparttable}{\vskip10pt}{}{}
\makeatletter
\patchcmd{\TPT@doparanotes}
  {\TPT@hsize}
  {\TPT@hsize\footnotesize}
  {}
  {}
\makeatother

\journal{Advances in Space Research}

\begin{document}

\verso{C.J.K Larkin \textit{et al.}}

\begin{frontmatter}

\title{M$^5$ --- Mars Magnetospheric Multipoint Measurement Mission: \newline A multi-spacecraft plasma physics mission to Mars}

\author[1,2,18,21]{Cormac J. K. \snm{Larkin}\corref{cor1}}
\cortext[cor1]{Corresponding author: 
   Tel.: +49 6221 54-1884;  
  }
  \ead{cormac.larkin@uni-heidelberg.de}


\author[3]{Ville \snm{Lund\'{e}n}}
\author[4]{Leonard \snm{Schulz}}

\author[5]{Markus \snm{Baumgartner-Steinleitner}}
\author[6]{Marianne \snm{Brekkum}}

\author[8]{Adam \snm{Cegla}}
\author[9,20]{Pietro \snm{Dazzi}}
\author[10]{Alessia \snm{De Iuliis}}
\author[11]{Jonas \snm{Gesch}}
\author[12]{Sofia \snm{Lennerstrand}}
\author[13]{Sara \snm{Nesbit-\"{O}stman}}
\author[7]{Vasco D. C. \snm{Pires}}
\author[14]{In\'{e}s \snm{Terraza Palanca}}
\author[15,19]{Daniel \snm{Teubenbacher}}

\author[16]{Florine \snm{Enengl}}
\author[17]{Marcus \snm{Hallmann}}

\address[1]{Zentrum f\"{u}r Astronomie der Universit\"{a}t Heidelberg, Astronomisches Rechen-Institut, M\"{o}nchhofstr. 12-14, 69120 Heidelberg, Germany}
\address[2]{Max-Planck-Institut f\"{u}r Kernphysik, Saupfercheckweg 1, 69117 Heidelberg, Germany}
\address[18]{Max-Planck-Institut f\"{u}r Astronomie, K\"{o}nigstuhl 17, D-69117 Heidelberg, Germany}
\address[21]{Kapteyn Astronomical Institute, University of Groningen, Landleven 12, 9747 AD Groningen, the Netherlands}

\address[3]{Department of Electronics and Nanotechnology, School of Electrical Engineering, Aalto University, Maarintie 8, 02150 Espoo, Finland}
\address[4]{Institute of Geophysics and Extraterrestrial Physics, Technische Universit\"{a}t Braunschweig, Mendelssohnstr. 3, 38106 Braunschweig, Germany}
\address[5]{Institute of Theoretical and Computational Physics, Graz University of Technology, 8010 Graz, Austria}
\address[6]{University of South-Eastern Norway, Raveien 215, 3184 Borre, Norway}

\address[8]{Institute of Geodesy and Geoinformatics, Wroc\l{}aw University of Environmental and Life Sciences, Grunwaldzka 53, 50-375 Wroc\l{}aw, Poland}
\address[9]{Laboratoire de Physique et Chimie de l’Environnement et de l’Espace (LPC2E), CNRS, Université d’Orl\'eans, Orl\'eans, France}
\address[20]{LESIA, Observatoire de Paris, PSL Research University, CNRS, Sorbonne Universit\'e, UPMC, Universit\'e Paris Diderot, Sorbonne Paris Cit\'e, Meudon, France}
\address[10]{Politecnico di Torino, Corso Duca degli Abruzzi, 24, 10129 Torino, Italy}
\address[11]{Institute of Optical Sensor Systems, Deutsches Zentrum f\"{u}r Luft- und Raumfahrt e.V., Rutherfordstr. 2,  12489 Berlin, Germany}
\address[12]{Department of Systems and Space Engineering, Lule\r{a} University of Technology, SE-971 87 Lule\r{a}, Sweden}
\address[13]{Department of Physics, Ume\r{a} University, SE-901 87 Ume\r{a}, Sweden}
\address[7]{DEMec, Faculty of Engineering, University of Porto, R. Dr. Roberto Frias 400, 4200-465 Porto, Portugal}
\address[14]{Facultat de F\'{i}sica i Qu\'{i}mica, Universitat de Barcelona, Carrer de Mart\'{i} i Franqu\`{e}s, 1, 11, 08028 Barcelona, Spain}
\address[15]{Space Research Institute, Austrian Academy of Sciences, Schmiedlstrasse 6, 8042 Graz, Austria}
\address[19]{Institute of Physics, University of Graz, Universitätsplatz 5, 8010 Graz, Austria}
\address[16]{Department of Physics, University of Oslo, Problemveien 7, 0315 Oslo}
\address[17]{German Aerospace Center (DLR), Institute of Space Systems, Robert-Hooke-Str. 7, 28359 Bremen}

\received{1 May 2013}
\finalform{10 May 2013}
\accepted{13 May 2013}
\availableonline{15 May 2013}
\communicated{S. Sarkar}

\begin{abstract}
Mars, lacking an intrinsic dynamo, is an ideal laboratory to comparatively study induced magnetospheres, which can be found in other terrestrial bodies as well as comets. Additionally, Mars is of particular interest to further exploration due to its loss of habitability by atmospheric escape and possible future human exploration. In this context, we propose the \textit{Mars Magnetospheric Multipoint Measurement Mission} (M$^5$), a multi-spacecraft mission to study the dynamics and energy transport of the Martian induced magnetosphere comprehensively. Particular focus is dedicated to the largely unexplored magnetotail region, where signatures of magnetic reconnection have been found. Furthermore, a reliable knowledge of the upstream solar wind conditions is needed to study the dynamics of the Martian magnetosphere, especially the different dayside boundary regions but also for energy transport phenomena like the current system and plasma waves. This will aid the study of atmospheric escape processes of planets with induced magnetospheres. In order to resolve the three-dimensional structures varying both in time and space, multi-point measurements are required. Thus, M$^5$ is a five spacecraft mission, with one solar wind monitor orbiting Mars in a circular orbit at 5 Martian radii, and four smaller spacecraft in a tetrahedral configuration orbiting Mars in an elliptical orbit, spanning the far magnetotail up to 6 Mars radii with a periapsis within the Martian magnetosphere of 1.8 Mars radii. We not only present a detailed assessment of the scientific need for such a mission but also show the resulting mission and spacecraft design taking into account all aspects of the mission requirements and constraints such as mass, power, and link budgets. Additionally, different aspects of the mission programmatics like a possible mission timeline, cost estimates, or public outreach are shown. The common requirements for acceptance for an ESA mission are considered. The mission outlined in this paper was developed during the Alpbach Summer School 2022 on the topic of ``Comparative Plasma Physics in the Universe''.

\end{abstract}


\end{frontmatter}


\section{Introduction}
\label{sec1}

\begin{figure*}[!b]
    \centering
    \includegraphics[width=1\linewidth]{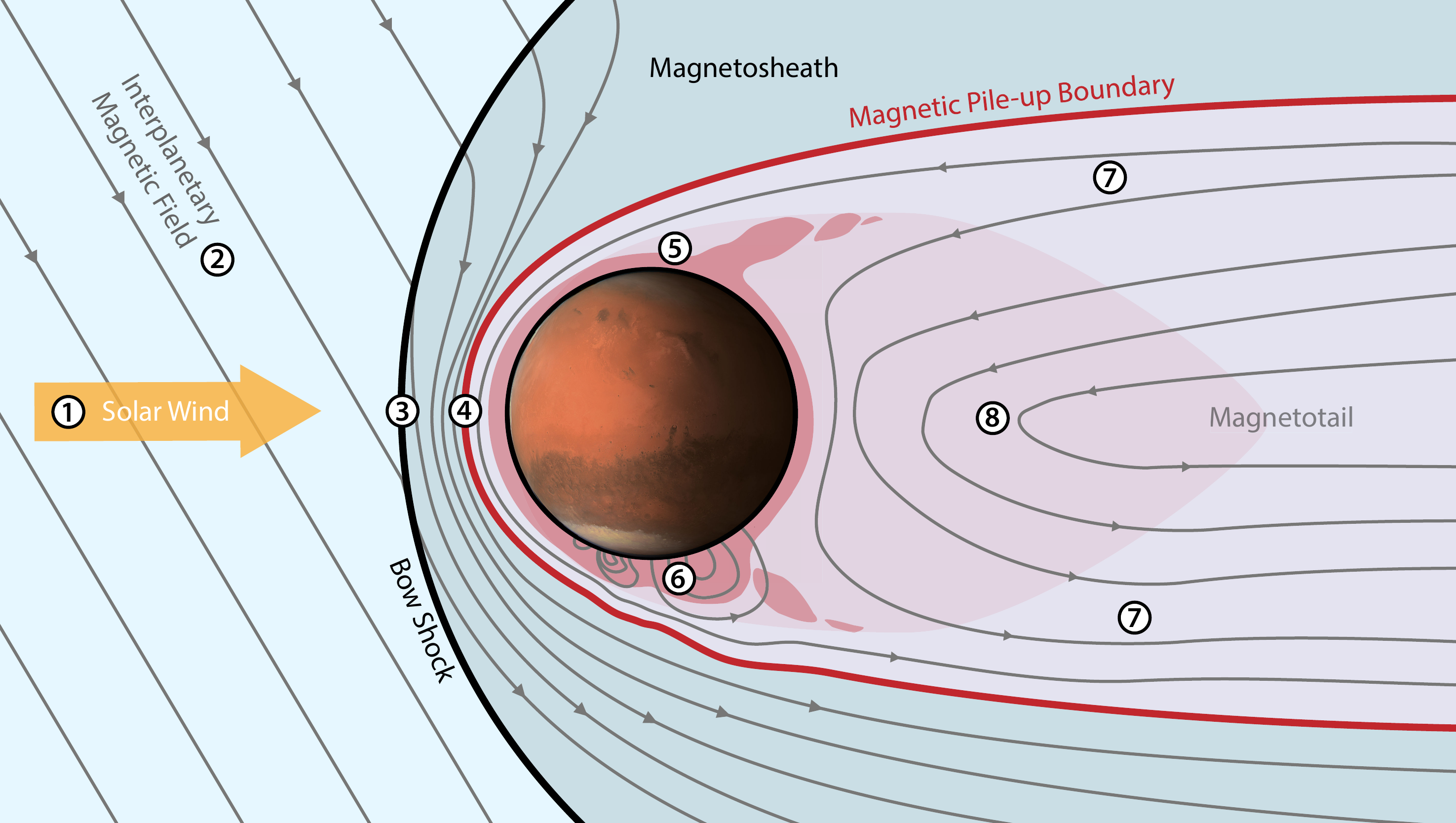}
    \caption{Overview of the Martian induced magnetosphere. The Interplanetary Magnetic Field (IMF) is draped around the planet, forming boundary regions and a highly dynamical magnetotail that is yet to be studied in detail. The numbers indicate the different plasma zones addressed in the text. 1. Solar wind, 2. IMF, 3. Sub-solar point of the bow shock, 4. Sub-solar point of the magnetic pile-up boundary, 5. Ionosphere, 6. Crustal field, 7. Lobes of the magnetotail, 8. Plasma sheet of the magnetotail.}
    \label{fig:magnetosphere}
\end{figure*}

Among the planets in the solar system, Earth, Mercury, and the gas giants possess a global intrinsic magnetic field due to an active internal dynamo process. This is the dominant driver in the deflection and thermalization of the solar wind plasma. The region where the solar wind dynamic is influenced by the planet's magnetic field is called the magnetosphere. However, other planets such as Mars \citep{dubinin2015} and large solar system bodies like the Moon do not show such a dynamo and therefore lack a global intrinsic magnetic field. These bodies can still have local intrinsic magnetic fields --- \textcolor{red}{Mars possesses strong magnetic anomalies (crustal fields) of up to \SI{700}{\nano\tesla} at 200 km altitude, which are at least one order of magnitude more intense than the crustal fields on Earth  \citep{Langlais_etal2019}} --- but in general, the large scale interaction with the solar wind of such systems is much different. For Mars, the direct interaction with the upper-atmosphere generates the so called induced magnetosphere \citep{sanchez2021}. The different regions of the Martian magnetosphere are presented in  \autoref{fig:magnetosphere}. Referring to the numbers in the figure, the Interplanetary Magnetic Field (IMF, 2) draped around the planet interacts with the solar wind (1), forming a bow shock (BS, 3) and a magnetic pileup boundary (MPB, 4), resembling the magnetopause at Earth, as dayside boundary regions \citep{art_Trotignon_2006} above the ionosphere (5). On the nightside, there is the magnetotail with its two lobes (7) that are separated by a plasma sheet (8), directed in opposite directions \citep{eastwood2008}. Due to the induced character of the magnetosphere, the average sub-solar bow shock distance (3) at 0.63 planetary radii from the surface \citep{art_Trotignon_2006} is much shorter than compared to e.\,g. Earth at about 13 Earth radii. \textcolor{red}{The crustal fields (6) of Mars can standoff the solar wind \citep{Brain_etal2003}}.

It is believed that Mars used to be more Earth-like, with a wetter and warmer climate. For this to have been the case, the atmosphere must have been denser than at present \citep{jakosky_2017}. Today, this is no longer the case, and in order to answer the question of how Mars became less habitable, we must investigate how the atmosphere was lost over time. \textcolor{red}{This investigation starts with studying atmospheric loss in the present, from which one can then attempt to extrapolate the loss rates back in time. The absence of a global magnetic field makes the process different to that at Earth, specifically in terms of ion loss. Whether the presence of a global magnetic field protects the atmosphere from ion loss is up for debate, with some evidence suggesting that it actually increases ion escape \citep{ Gunell_etal2018, Sakata_etal2020, Ramstad_etal2021}.  The presence of crustal fields at Mars and the ensuing hybrid nature of its magnetosphere adds further complexity, with the crustal fields both inhibiting and enabling ion loss \citep{Brain_etal2010, Ma_etal2014, Fang_etal2017, Dubinin_etal2020}. }
\textcolor{red}{Today, ion loss is a small part of the atmospheric loss at Mars, but may have been more significant in the past \citep{Jakosky_etal2018}. Ions escape through a multitude of processes, many of  which have been mapped by the MAVEN mission \citep[][and references therein]{Jakosky_etal2018}. What is missing currently is consistent solar wind monitoring combined with simultaneous in-situ measurements of the Martian magnetosphere, to enable studies of how these processes are affected by different solar wind conditions and by solar activity. By gaining a deeper understanding of ion loss dependence on different solar wind conditions and solar activity, further extrapolations can be made on how atmospheric escape has changed through time.}

Additionally, the knowledge of space weather at Mars is an important driver for future exploration of Mars. \textcolor{black}{Solar events, like interplanetary coronal mass ejections, Solar Energetic Particles, fast stream, etc., cause a high variability in the Martian magnetosphere \citep{hanaoka2023origin}. This poses a threat to spacecraft and space infrastructure flying within the induced magnetosphere \citep{Hassler17_SpaceWether}, with possible catastrophic consequences \citep{marusek2007solar}.} Moreover, astronaut safety in the future manned exploration of Mars could be jeopardized if the conditions at Mars are not known in detail \citep{art_Cucinotta_2013}. Therefore, near-continuous observations of the solar wind conditions at Mars are needed in order to both determine the average and extreme space weather conditions and determine their influence on the Martian magnetospheric system. Furthermore, a dedicated Martian solar wind observatory not only extends the ``orchestra'' of solar wind monitors, \textcolor{black}{but also could aid in the study of the evolution of solar events.}

\textcolor{red}{Mars offers the opportunity to study an induced magnetosphere in greater detail. Due to Mars’ proximity to Earth within the solar system, it can feasibly be reached by in-situ instrumentation. Not only is it a representative example of a solar system induced magnetosphere (like Venus), but also relevant to studies of comets and active asteroids \citep{goetz19}.} Furthermore, if unique characteristic properties of such magnetospheric systems are identified, these could have implications for the characterization of exoplanetary plasma environments \citep{exoplanets_2020}. 

\textcolor{black}{Changes in the IMF components induce a reorientation of the tail \citep{DiBraccio_2017}, which is characteristic of this variability.}
In order to separate temporal and spatial variations of these moving or flapping structures in the tail, simultaneous multi-point measurements are needed. Despite comprehensive studies of the Martian environment of previous missions, the far tail region has never been characterized in detail by in-situ measurements. A current open question is whether magnetic reconnection of the IMF occurs in the far tail at Mars, and if so, to what extent.

Magnetic reconnection is a fundamental plasma process where magnetic energy is converted to kinetic energy. It has been studied at Earth with formation missions like Cluster and the Magnetospheric Multiscale (MMS) mission. Similar processes occur on other magnetized and unmagnetized planets. On Mars, both measurements \citep{Harada_2015, Wang_2021}, and simulations \citep{Ma_2018} suggest that reconnection occurs on the nightside, playing a role in the dynamics of the magnetotail influencing ion flow velocities with possible effects on atmospheric escape.

Reconnection is not the only physical process of interest that takes place in the magnetotail. The magnetotail is one of the main paths for planetary ions to escape from the Martian atmosphere \citep{Dibraccio_etal2015, Brain_etal2015, Dong_etal2015, Dubinin_etal2017, Lin_2021, Curry_etal2022}. Therefore, a mapping of the properties of the Martian magnetotail complements ongoing studies of this important process and will allow a more complete assessment of balancing terms of atmosphere system in- and outflow. This is \textcolor{black}{crucial} for the understanding of how habitability of Mars has changed over time. 

Moving from the Martian nightside to the dayside, \textcolor{black}{important} features of the induced magnetosphere are the BS and MPB. MAVEN \citep[e.g.][]{Jakosky2015} has \textcolor{black}{observed} this region, showing a strong variation of the position of both BS and MPB \citep{article_Matsunaga_2017}. However, a systematic characterization of their variability depending on solar wind conditions is lacking. Knowledge of the dependency of the system's short-term evolution on solar wind conditions --- especially for solar high-energy events --- is imperative for spacecraft and astronaut safety.

Energy transfer and transport, especially on global and ion-scales, is another important aspect of the characterization of the Martian magnetospheric system, which will help in understanding the complete picture of the evolution of the atmosphere. One of the ways to transport energy is by currents. A year-average picture of the Martian current system has been acquired by MAVEN \citep{article_Ramstad_2020}, but a detailed, time-varying characterization is lacking. To measure the instantaneous current, a tetrahedral multi-spacecraft configuration is needed, in which methods such as the curlometer technique can be used, as it has been done at Earth for Cluster \citep{dunlop2021curl}. This would allow the measuring of transient currents, which are lost in the process of averaging. Furthermore, by having a solar wind monitor, the response of the currents to changing solar wind conditions can be investigated. 

Another way of transferring energy is through plasma waves, which are important to study due to their ability to accelerate and scatter particles, which can lead to the escape of particles from the atmosphere. Many waves around Mars have been identified, such as Whistler waves, Proton Cyclotron waves and Magnetosonic waves \citep{article_Yadav_2021, article_Brain_2002}. Other waves such as Ion Acoustic waves and Lower Hybrid waves are predicted to exist in the Mars ionosphere, but have yet to be detected \citep{article_Yadav_2021}. The detection of the latter could explain some of the loss of particles from Mars outer ionosphere through particle acceleration. In order to fully characterize these waves, temporal and spatial variations would need to be resolved and separated, which requires a tetrahedron formation of spacecraft \citep{karlsson04,Narita2010}. 

In order to allow for the separation of spatial and temporal variations of 3D plasma structures, again a four-spacecraft tetrahedron constellation is needed. This has been demonstrated by the Cluster mission at Earth \citep{escoubet2021cluster}. This mission allows the characterization of the time variation of the dayside boundaries and simultaneously determine their 3D spatial extent. Additionally, currents on above-ion-scales were detected by Cluster using the curlometer technique \citep{dunlop2021curl}, as well as waves and turbulence with the wave-telescope technique \citep{art_Narita_2022} which are techniques only possible using four-point measurements. 

\textcolor{black}{Other missions at Earth have demonstrated how many important results can be obtained with a multi-spacecraft mission for space weather studies. The THEMIS mission, launched in 2007 and including five satellites, is designed to study space weather phenomena \citep{angelopoulos2009themis,sibeck2008themis,mcfadden2009themis}. THEMIS also allows for the important study of Earth's boundary regions, characterizing the current sheet thickness, motion and current density of the magnetopause \citep[e.\,g.][]{haaland2019characteristics}, amongst many other important results (\citealt{pvrech2008response,li2009cold, artemyev2020ionospheric} and more). Another successful multi-spacecraft mission is MMS, a four-spacecraft plasma research mission dedicated to characterizing reconnection \citep{burch2016}. MMS was the first spacecraft able to measure reconnection on electron scales, which was then studied by \cite{burch2016magnetic,hesse2016electron, shay2016kinetic} and many more.} All this shows the success and need for a four-spacecraft constellation to study a planetary magnetospheric system comprehensively.

\begin{table*}[t]
\caption{Scientific questions and objectives of the M$^5$ mission. The specific regions, that are referred to by the scientific objectives are given by numbers in parenthesis, corresponding to the regions specified in Figure \ref{fig:magnetosphere}.}
    \centering
        \begin{tabular}{*{15}{p{50mm}|*{15}{p{120mm}}}}
        \textbf{Primary scientific question} & \textbf{Primary scientific objectives} \\ \hline
        \textbf{Q1}: How do the Martian magnetospheric system’s structure and dynamics depend on solar wind conditions? 
        & 
        O1.1 (1, 3, 4): What are the dynamics and orientation of boundary regions, with particular interest for their dependence upon solar wind conditions? \newline
        O1.2 (1, 7, 8): What is the structure of the Martian magnetotail on different scales, with particular interest for its dependence upon solar wind conditions? \newline
        O1.3 (1, 3, 4, 5, 7, 8): What is the dynamical structure of the current system in the Martian magnetosphere, with particular interest for its dependence upon solar wind conditions? 
        \\ \hline
        \textbf{Q2}: How is energy transported within the Martian magnetospheric system on ion scales and above?
        & 
        O2.1 (7, 8): Is magnetic reconnection observed in the magnetosphere tail, and if so, where and how? \newline
        O2.2 (3, 4): What are the direction and temporal evolution of low frequency plasma waves?
        \\ \hline \hline
        \textbf{Secondary scientific question} & \textbf{Secondary scientific objectives} \\ \hline
        \textbf{Q3}: How does the solar wind propagate through the solar system?
        & 
        O3.1 (1): What are the temporal variations of the upstream solar wind conditions at Mars?
        \\ \hline
        \textbf{Q4}: Excluding magnetic reconnection, are there other processes driving the energy transport at the Martian magnetotail?
        &
        O4.1 (7, 8): Are other energy transport processes observed at the Martian magnetotail that exhibit signatures different to magnetic reconnection?
        \\
        \end{tabular}
    \label{tab_scienceobjs}
\end{table*}\vspace{0mm}

In the last decades, multiple missions have targeted Mars, tackling diverse science topics like the search for water and bio-signatures and the exploration of Mars' surface. The ongoing missions Mars Express \citep{chicarro2004mars} and MAVEN \citep{Jakosky2015} \textcolor{red}{have greatly contributed to our understanding of the Martian atmospheric composition, evolution and circulation}. They are also equipped with plasma instrument suites, however are limited as for example Mars Express lacks a magnetometer. Additionally, the scientific output on the Martian magnetosphere is limited due to the lack of additional orbiters which would allow the observation of temporal and spatial variations. Moreover, there is currently no dedicated solar wind monitor at Mars, which is needed to investigate the variability of the magnetosphere depending on solar wind conditions. 
\textcolor{red}{}

The upcoming mission \textit{Escape and Plasma Acceleration and Dynamics Explorers} (EscaPADE) --- \textcolor{red}{scheduled to launch in August 2024, arrive at Mars in September 2025, and officially start its science campaign March 2026} --- will study the flow of both energy and ions in and out of the Martian atmosphere \citep{Lillis_2022}. \textcolor{red}{It will be the first twin-spacecraft space plasma mission beyond Earth's orbit. EscaPADE will have two consecutive science campaigns, the first a six month string-of-pearls configuration, and the second being separate orbits where the planes precess differentially. Its capacity to produce dual-point measurements will enable great scientific progress on the Martian plasma environment, upon which a multi-point mission could build. For instance, a tetrahedron configuration would uniquely enable the three-dimensional study of phenomena such as currents, waves and reconnection using known multi-spacecraft analysis techniques. By combining this with a solar wind monitor, the impact on these from varying solar wind conditions and solar activity could be studied.} \textit{Mars Magnetosphere ATmosphere Ionosphere and Surface SciencE} (M-MATISSE) is a mission currently being studied for the ESA M7 call aiming to characterise the region between the Martian upper atmosphere and the outer magnetosphere, and to study how surface processes are affected by space weather \citep{M-Matisse_2022}. \textcolor{red}{Further upcoming missions to Mars include the Japanese Mars Moons Explorer (MMX) \citep{Kuramoto_etal2022} mission which will be able to make magnetic field and suprathermal ion measurements including the solar wind, and the Tianwen-1 \citep{Zou_etal2021} mission which will have the capacity to measure the magnetic field and ions. Notably, DC electric field measurements were proposed as part of the MOSAIC 10-spacecraft constellation to study the Martian climate system from subsurface ice all the way to the solar wind \citep{lillis2021}. However, none of the plasma missions sent to Mars to date have been capable of measuring DC electric fields.}

Despite the considerable number of Martian exploration missions, there has been a paucity of plasma physics-focused missions in the past. Furthermore, both of the future dedicated plasma missions lack the capabilities to produce a complete and detailed picture of the structures and energy transport with both temporal and spatial dependencies in the whole Martian induced magnetospheric system as well as providing this information with dependency on precise upstream solar wind conditions.

All in all, the change of the magnetosphere with solar wind conditions and how energy is transferred across different scales --- both spatially and temporally --- remain to be fully understood. Additionally, the Martian magnetotail is still largely unexplored. This is reflected in the \textit{Voyage 2050 Senior Committee Report} \citep{voyage2050seniorcommittee_2021_voyage}, which was written to identify key science areas for ESA's science program during the period 2035-2050. Relevant key areas are ``Magnetospheric Systems'' (3.1.1) and ``Plasma Cross-scale Coupling'' (3.1.2). They state that, \textit{``important questions such as 'How is energy and matter transported in induced magnetospheres' still need to be answered by studying entire magnetospheres as complex systems''}. In this context, we propose the \textit{Mars Magnetospheric Multipoint Measurement Mission}, hereafter M$^5$, a 5-spacecraft mission to study the different regions of the Martian magnetosphere comprehensively, by using a four-spacecraft tetrahedron formation for in-situ measurements while monitoring the solar wind with an additional spacecraft.

This paper is organized as follows: in Section~\ref{sec:theme_and_questions}, the Scientific Objectives and Questions, derived from the above shown open research areas are given. With that, measurement requirements for different physical quantities to be measured at Mars are specified. Subsequently, the mission profile is described in Section~\ref{sec:mission_profile}, with the required scientific payload following in Section~\ref{sec:payload}. In Section~\ref{sec:mission_design}, all technical aspects of the proposed mission are assessed in detail. Finally, programmatics are addressed in Section~\ref{sec:programmatics} followed by a general conclusion (Section~\ref{sec:conclusion}).

\section{Scientific Questions and Measurement Requirements} \label{sec:theme_and_questions}

\begin{table*}[h]
 \caption{Scientific objective addressed by each instrument used by the M$^5$ mission. A big dot $\bigcirc$ stands for the Solar Wind Orbiter (SWO) and a small dot $\bullet$ for an Magnetospheric Formation Orbiter (MFO).}
    \centering
\begin{tabular}{ c||c|c c c c c }
\hline
\begin{tabular}[c]{@{}c@{}}Science \\ question\end{tabular} & \begin{tabular}[c]{@{}c@{}}Science \\ objective\end{tabular} & 
\begin{tabular}[c]{@{}c@{}}DC Vector \\ magnetic field \end{tabular} & \begin{tabular}[c]{@{}c@{}}Ion \\ distribution function \end{tabular} & \begin{tabular}[c]{@{}c@{}}Electron \\ distribution function\end{tabular} & \begin{tabular}[c]{@{}c@{}}Density \\ temperature\end{tabular} & \begin{tabular}[c]{@{}c@{}}DC Vector \\ electric field\end{tabular} \\ \hline 

\begin{tabular}[c]{@{}c@{}} \\ \end{tabular} & \begin{tabular}[c]{@{}c@{}} \\ \end{tabular} & 
Magnetometer & \begin{tabular}[c]{@{}c@{}}Ion \\ spectrometer\end{tabular} & \begin{tabular}[c]{@{}c@{}}Electron \\ spectrometer\end{tabular} & \begin{tabular}[c]{@{}c@{}}Langmuir \\ probe\end{tabular} & \begin{tabular}[c]{@{}c@{}}Dipolar \\ antennas\end{tabular} \\ \hline 

Q1 & O1.1 & $\bigcirc$ $\bullet$ $\bullet$  $\bullet$ $\bullet$ & $\bigcirc$ $\bullet$ &   &   &   \\ \cline{2-7} 
   & O1.2 & $\bigcirc$ $\bullet$ $\bullet$  $\bullet$ $\bullet$ & $\bigcirc$ $\bullet$ & $\bigcirc$ & $\bullet$ $\bullet$ $\bullet$ $\bullet$ & \\ \cline{2-7} 
   & O1.3 & $\bigcirc$ $\bullet$ $\bullet$ $\bullet$ $\bullet$ & $\bigcirc$ & $\bigcirc$ & & \\ \hline
Q2 & O2.1 & $\bullet$ $\bullet$ & $\bullet$ $\bullet$ & & &  \\ \cline{2-7} 
   & O2.2 & $\bullet$ $\bullet$ $\bullet$ $\bullet$ & & & $\bullet$ $\bullet$ $\bullet$ $\bullet$ & $\bullet$ $\bullet$ $\bullet$ $\bullet$ \\ \hline
Q3 & O3.1 & $\bigcirc$ & $\bigcirc$ & $\bigcirc$ & & \\ \hline
Q4 & O4.1 &  $\bullet$ $\bullet$ & $\bullet$ $\bullet$ & $\bullet$ $\bullet$ & & $\bullet$ $\bullet$ \\ \hline
\end{tabular}
    \label{tab_instrumentation}
\end{table*}\vspace{-1mm}

In order to structure the different regions and physical phenomena and make them more approachable from an instrument point of view, we define a broad scientific theme for the M$^5$ mission:\textcolor{black}{\textit{``To understand how the variable solar wind conditions influence the dynamics and energy transport of the Martian induced magnetosphere."}}

From that, two primary scientific questions are derived, which are then segmented into scientific objectives. This hierarchy is shown in Table~\ref{tab_scienceobjs}, including reference to the regions of interest shown in Figure~\ref{fig:magnetosphere}. 

The first primary scientific question ($\textbf{Q1}$) focuses on the dependency of the Martian magnetosphere on solar wind conditions. The second question ($\textbf{Q2}$) relates to energy transport in the Martian magnetosphere. In addition to these two primary scientific questions, M$^5$ will be able to tackle two other secondary scientific questions. The third question ($\textbf{Q3}$) concentrates on the propagation of the solar wind in the solar system. The fourth question ($\textbf{Q4}$) is related to the possibility that reconnection in the Martian magnetotail is not the only process driving energy transport.

The respective scientific objectives allow for the definition of measurement requirements by using a traceability matrix. ~\autoref{tab_instrumentation} shows the required measurement quantities for instruments on each spacecraft respectively, both on the \textit{Solar Wind Observatory} (SWO) and the four \textit{Magnetospheric Formation Orbiters} (MFO) constituting a tetrahedron constellation. The requirements were derived from each of the measurement regions, physical quantities, timing constraints, and specific measurement needs (e.g. range and accuracy) in question. The typical parameters that are expected to be observed by the M$^5$ missions are derived by previous in-situ measurements \citep{ Nilsson_2012, Holmberg_2019, Ergun_2021}.

The requirements for magnetic field, ion distribution functions, electron distributions functions, and electric field measurements are detailed in \autoref{tab:magnetometer}, \autoref{tab:ion_reqs}, \autoref{tab:electron_reqs}, and \autoref{tab:electric_field_reqs} respectively. Based on the measurement requirements, corresponding heritage instruments or instrument options have been selected and are presented in Section~\ref{sec:payload}.

\begin{table}[h]
\centering
\caption{Magnetic field measurement requirements}  \label{tab:magnetometer}
\begin{tabular}{ c c c }
	\hline
Requirement & In Magnetosphere & In Solar Wind \\
	\hline 
Absolute range & \SI{3000}{\nano\tesla} & \SI{500}{\nano\tesla}   \\ 	
Absolute accuracy \textcolor{red}{(per axis)} & \SI{0.5}{\nano\tesla} & \SI{0.5}{\nano\tesla} \\ 	
Temporal resolution  &  \SI{32}{sps} & \SI{32}{sps} \\  \hline
\end{tabular} 
\end{table}

\begin{table}[h]
\centering
\caption{Ion moments measurement requirements}  \label{tab:ion_reqs}
\begin{tabular}{ c c c }
	\hline
Requirement & In Magnetosphere & In Solar Wind \\
	\hline 
Energy range & \SI{1}{eV}--\SI{30}{keV} & \SI{10}{eV}--\SI{25}{keV}  \\ 
Energy resolution & \SI{25}{\percent} & \SI{25}{\percent} \\
\textcolor{red}{\begin{tabular}[c]{@{}c@{}}Differential energy flux range\end{tabular}} & \textcolor{red}{\begin{tabular}[c]{@{}c@{}}$10^4$--$10^{10}$\\$\mathrm{eV/(eV\,cm^2\,s\,sr)}$\end{tabular}} & \textcolor{red}{\begin{tabular}[c]{@{}c@{}}$10^4$--$10^{10}$\\$\mathrm{eV/(eV\,cm^2\,s\,sr)}$\end{tabular}} \\
Temporal resolution  &  \SI{5}{\second} & \SI{5}{\second} \\  
FoV  &  \ang{360}~$\times$~\ang{90} & \ang{180}~$\times$~\ang{40} \\  
Ions to detect  & \begin{tabular}[c]{@{}c@{}}H+, He++,\\higher mass\end{tabular}  & \begin{tabular}[c]{@{}c@{}}H+, He++,\\higher mass\end{tabular} \\  \hline
\end{tabular} 
\end{table}

\begin{table}[h]
\centering
\caption{Electron moments measurement requirements}  \label{tab:electron_reqs}
\begin{tabular}{ c c c }
	\hline
Requirement & In Magnetosphere & In Solar Wind \\
	\hline 
Energy range & \SI{50}{eV}--\SI{10}{keV} & \SI{10}{eV}--\SI{5}{keV}  \\ 	
Energy resolution & \SI{25}{\percent} & \SI{25}{\percent} \\
\textcolor{red}{\begin{tabular}[c]{@{}c@{}}Differential energy flux range\end{tabular}} & \textcolor{red}{\begin{tabular}[c]{@{}c@{}}$10^4$--$10^{10}$\\$\mathrm{eV/(eV\,cm^2\,s\,sr)}$\end{tabular}} & \textcolor{red}{\begin{tabular}[c]{@{}c@{}}$10^4$--$10^{10}$\\$\mathrm{eV/(eV\,cm^2\,s\,sr)}$\end{tabular}} \\
Temporal resolution  &  \SI{5}{\second} & \SI{5}{\second} \\
FoV  &  \ang{360}~$\times$~\ang{120} & \ang{180}~$\times$~\ang{40} \\  \hline
\end{tabular} 
\end{table}

\begin{table}[h]
\centering
\caption{Electric field measurement requirements}  \label{tab:electric_field_reqs}
\begin{tabular}{ c c c }
	\hline
Requirement & In Magnetosphere & In Solar Wind \\
	\hline 
Absolute range & ±\SI{300}{\milli\volt/\meter} & --   \\ 	
Accuracy & \SI{1}{\milli\volt/\meter} or \SI{10}{\percent}& -- \\ 	
Temporal resolution  &  \SI{1}{Hz}--\SI{200}{Hz} & -- \\  \hline
\end{tabular} 
\end{table}

\section{Mission Profile}\label{sec:mission_profile}

To answer the science questions and objectives stated in Table~\ref{tab_scienceobjs}, the M$^5$  mission requires a tetrahedral formation of four spacecraft. This allows the resolution of both spatial and temporal variations, as well as a three-dimensional mapping of the boundary regions, even when the location, velocity, and orientation of the boundary are unknown. This will result for example in the ability to take into account nonuniform conditions such as ripples and reformation, as has been done with Cluster.
The same applies to the largely unexplored \textcolor{red}{far} magnetotail. In addition, such a constellation enables the mapping of currents in the magnetosphere, using the curlometer technique \citep{art_Dunlop_1988} to derive currents from magnetic field measurements. Furthermore, it will be used for measurements of wave direction and time dependency using the wave telescope technique \citep{Motschmann_1996}. Finally, multiple spacecraft are needed to determine origin regions of magnetic reconnection by observing ion outflow.
Spacecraft separation distances on and above ion scales are required to observe all the mentioned phenomena. Ion scales at Mars range from the proton gyroradius in the near tail on the magnitude of 100\,km, to around 750\,km maximum in the magnetosheath \citep{Nilsson_2012}. In addition, an active solar wind monitor is needed to provide necessary simultaneous information about the solar wind conditions.

\begin{figure}[h]
\centering
\includegraphics[width=0.95\linewidth]{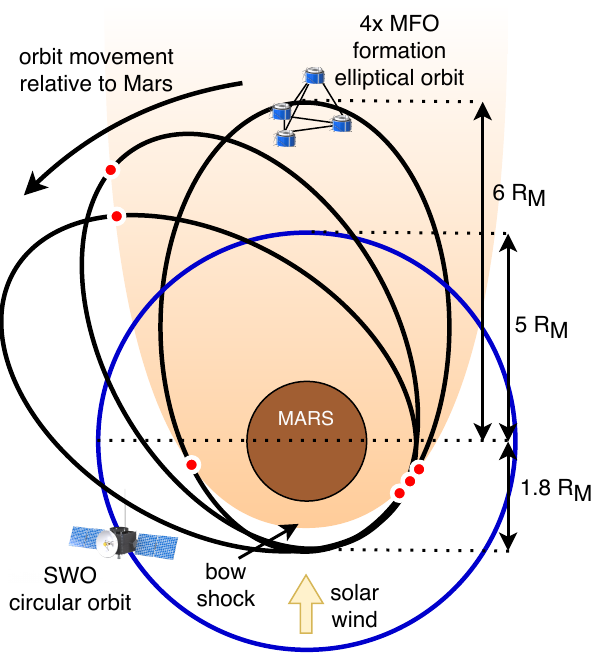}
\caption{\textcolor{black}{Final orbit configuration of MFOs and SWO at Mars. Due to orbit precession, the orbit of the MFOs will move relative to the Martian reference frame during the Martian year "sweeping" over regions of interest (e.g. boundary crossings marked with red dots). } }
\label{fig:orbit}
\end{figure}

Therefore, we propose a five spacecraft mission. Four identical MFOs will be placed in an elliptic orbit in a tetrahedral cartwheel helix formation\textcolor{black}{. The orbit is chosen in such a way that throughout a whole Martian year, the spacecraft spend a significant time in the far magnetotail. \textcolor{red}{In its initial configuration, the dayside periapsis of the orbit is chosen just slightly larger than the expected bow shock stand-off distance, while the apoapsis is in the far magnetotail. This guarantees a sufficient number of boundary crossings.} Orbit precession will gradually bring the apoapsis towards the dayside, thus allowing for a scanning of different boundary locations as well as the near tail region, as the periapsis moves to the nightside. A schematic of the orbits and the precession effects is shown in \autoref{fig:orbit}. Combined with a substantial orbit inclination, this way the MFOs will cover large portions of the Martian magnetotail and the boundary regions as well as the magnetosheath, addressing all primary science objectives of the mission. On-board fuel will allow for adjusting the tetrahedral configuration throughout the mission duration. Details of the final orbit configuration are given in \autoref{sec:orbit}.}

\begin{figure*}[h!] 
  \centering
  \begin{subfigure}[b]{0.48\textwidth}
    \includegraphics[width=8cm]{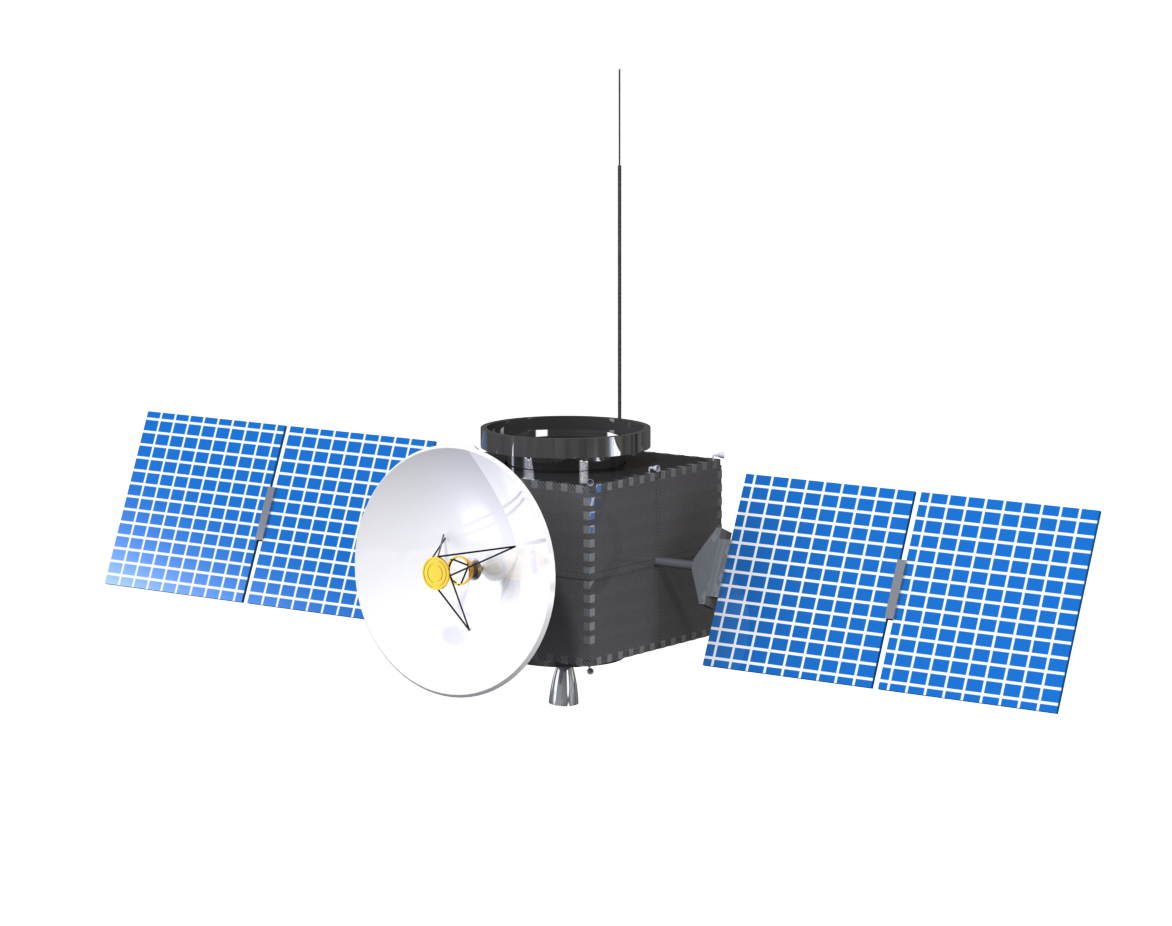}
    \caption{The Solar Wind Observatory.} 
  \end{subfigure}
  \hfill
  \begin{subfigure}[b]{0.48\textwidth}
    \includegraphics[width=8cm]{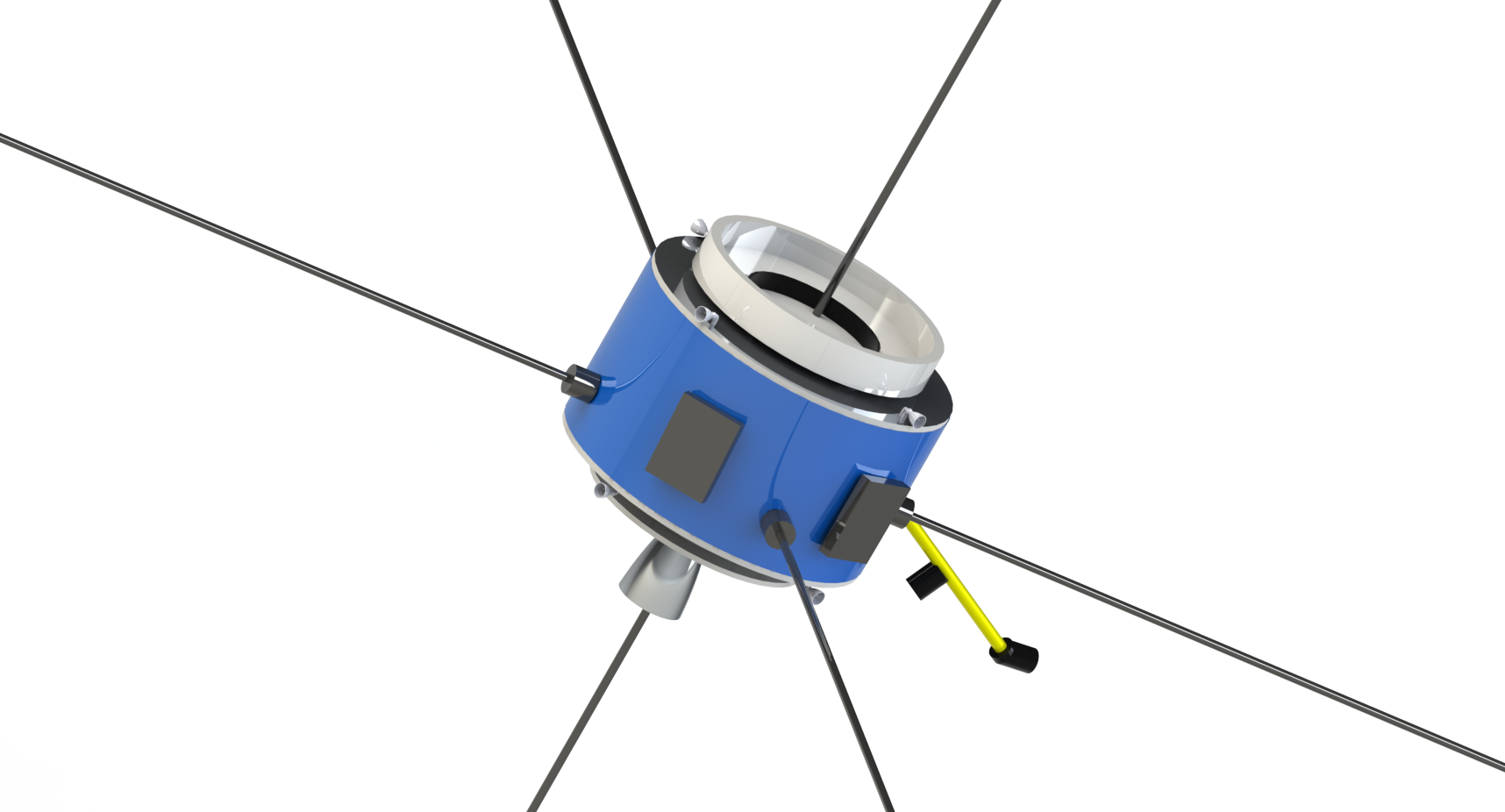}
    \caption{A Magnetospheric Formation Orbiter.} 
  \end{subfigure}
  \hfill
  \caption{Three-dimensional rendering of the two spacecraft types forming the M$^5$ mission.}\label{Fig:M5 mission}
\end{figure*}

The fifth spacecraft, the \textit{Solar Wind Observatory} (SWO), targets a circular orbit around Mars \textcolor{black}{(see \autoref{fig:orbit})}. The SWO will characterize the solar wind properties around Mars during the whole Martian year, thus addressing the secondary science question Q3, which supports addressing the primary science question Q1. As a result of the chosen orbit the SWO will spend a part of its orbit in the magnetotail, covering a region similar to the one explored by MAVEN. Furthermore, it acts as a data relay for the MFOs to Earth. \autoref{Fig:M5 mission} \textcolor{black}{shows both the SWO and one MFO spacecraft} in their final configuration at Mars.

\section{Payloads}\label{sec:payload}
In this section, we provide an overview of the proposed instruments for the M$^5$ mission, in terms of the heritage instruments they are based on. \textcolor{black}{The estimated resources required by the payloads are collected into \autoref{tab:payload_resources} at the end of this section.} \textcolor{red}{Other, complementary instrumentation not considered here is discussed in \autoref{section:costs}.}

\subsection{Flux Gate Magnetometer (FGM)} \label{sec:fgm}
The magnetometers proposed for the mission are 3-axis fluxgate magnetometers with heritage from THEMIS \citep{FGM}. Each spacecraft will carry a pair of these magnetometers mounted on different locations of a deployable boom stretching \SI{5}{\meter} in length. One magnetometer will be located at the tip of the boom, whereas the other one halfway up the boom. This configuration allows for effective magnetic interference mitigation, as described in Section~\ref{sec:EMI}. 

\subsection{Ion spectrometers} \label{sec:ion_spectrometers}
The mission will utilize electrostatic analysers to measure the ion energy distribution function. The instrument placed on the SWO will be used as an ion energy spectrometer. A heritage instrument proposed for the task on the SWO is Solar Orbiter's SWA-HIS instrument \citep{SWA-HIS}. 

In contrast, the instrument on each of the MFOs will use magnets to act as a mass over charge spectrometer. As heritage, the Ion Composition Analyser (ICA) instrument from Rosetta  \citep{rosetta_ion,carr2007rpc} is considered a viable option. The ion mass spectrometer will measure the 3D distribution function of the ions to study how the particles interact with the solar wind.

\subsection{Electrostatic electron analyser} \label{sec:electrostatic-electron-analyzer}
In order to measure the electron composition of the plasma environment, an electrostatic electron analyser will be employed on all five spacecraft. The heritage of the instruments is from the SWA-EAS instrument of the Solar Orbiter \citep{SWA-HIS}. The solar wind electron analyser will measure the effects from the electron impact ionization from the solar wind as it encounters the Martian atmosphere. 

\subsection{Electric field instrument} \label{sec:electric-field-instrument}
In order to measure the 3D electric field vector of the plasma environment, each MFO will have an electric field instrument using 6 booms (4 wire booms, 2 telescopic booms). In addition, two orthogonal probes will have Langmuir probe capabilities. This will be used to measure the temperature and density of the plasma. The instrument proposed for the described purpose is the electric-field and wave instrument (EFW) that has  heritage from ESA's Cluster mission \citep{Cluster_EFW}. 


\begin{table*}[h!]
\centering
\textcolor{black}{
\caption{\textcolor{black}{Estimated resources required by the payloads of the mission. Each resource estimate is given for a single payload. Power consumption refers to the nominal power consumption when the payload is in use. The estimates are based on the heritage instrument considered in \cref{sec:fgm,sec:ion_spectrometers,sec:electrostatic-electron-analyzer,sec:electric-field-instrument}.}}  \label{tab:payload_resources}
\begin{tabular}{ c c c c c }
	\hline
Payload & Mass [kg] & Power [W] & Data Rate [kbps] & References \\
	\hline 
Fluxgate magnetometer & 0.4 & 0.8 & 6 & a \\
Ion spectrometer (SWO) & 2.2 & 2.8 & 6 & b, c \\  
Ion spectrometer (MFO) & 2.2 & 2.8 & 1 & c, d \\  
Electrostatic electron analyzer & 2.0 & 3.8 & 4 & b \\  
Electric field instrument \textcolor{red}{(incl. booms)} & 14  &  3.7 & 1.5 & e \\\hline
\end{tabular} 
\begin{tablenotes}
\item[a] \citet{FGM} 
\item[b] \citet{SWA-HIS}
\item[c] \citet{carr2007rpc}
\item[d] \citet{rosetta_ion}
\item[e] \citet{Cluster_EFW}
\end{tablenotes}
}
\end{table*}

\section{Mission Design}\label{sec:mission_design}
In the following, we will detail the technical aspects of the mission.

\subsection{Margin Philosophy} \label{sec:margins}
The margin philosophy adopted for the mission design is based on recommendations detailed by ESA \citep{esa_margins}. The applicable sections of the margin philosophy have been considered for all system budgets including mass, $\Delta V$, propellant, data, and link budgets, as well as the power and thermal budgets.

\subsection{Ground Segment}
For ground segment communications section, the ESA Deep Space Antennas network, which include the antennas located in Cebreros (Spain), Malargüe (Argentina) and New Norcia (Australia) will be used. Science operations will take place at the European Space Astronomy Centre (ESAC), close to Madrid. 

\subsection{Launch \& Propellant}
\begin{figure}[h]
\centering
\includegraphics[width=0.95\linewidth]{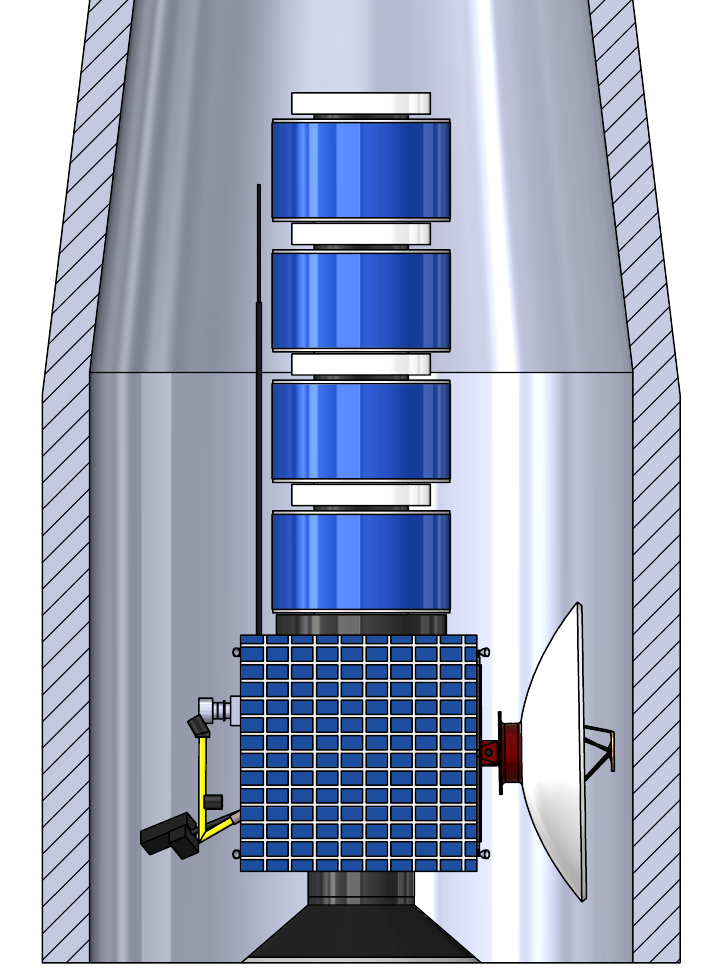}
\caption{Spacecraft in the launch configuration inside the Ariane's fairing.}
\label{fig:spacecraft_fairing}
\end{figure}

The M$^5$ mission is designed to be launched using an Ariane 64 launcher from Kourou, French Guiana. \autoref{fig:spacecraft_fairing} presents the M$^5$ mission spacecraft in the launch configuration inside the Ariane 64 fairing. After the launch, the five spacecraft will utilize thrusters with MMH/N204 bipropellant in order to perform the orbital and attitude maneuvers needed to reach and maintain the required orbits, stabilization, and attitude of the spacecraft. Helium pressurizing is used in order to maintain the operating pressures. Heritage thrusters from the ExoMars orbiter with a bi-propellant propulsion system \citep{pavon2012engineering} are proposed for the M$^5$ mission.

\subsection{Orbits \& Maneuvers} \label{sec:orbit}
\begin{figure*}[h]
\centering
\includegraphics[width=1.0\linewidth]{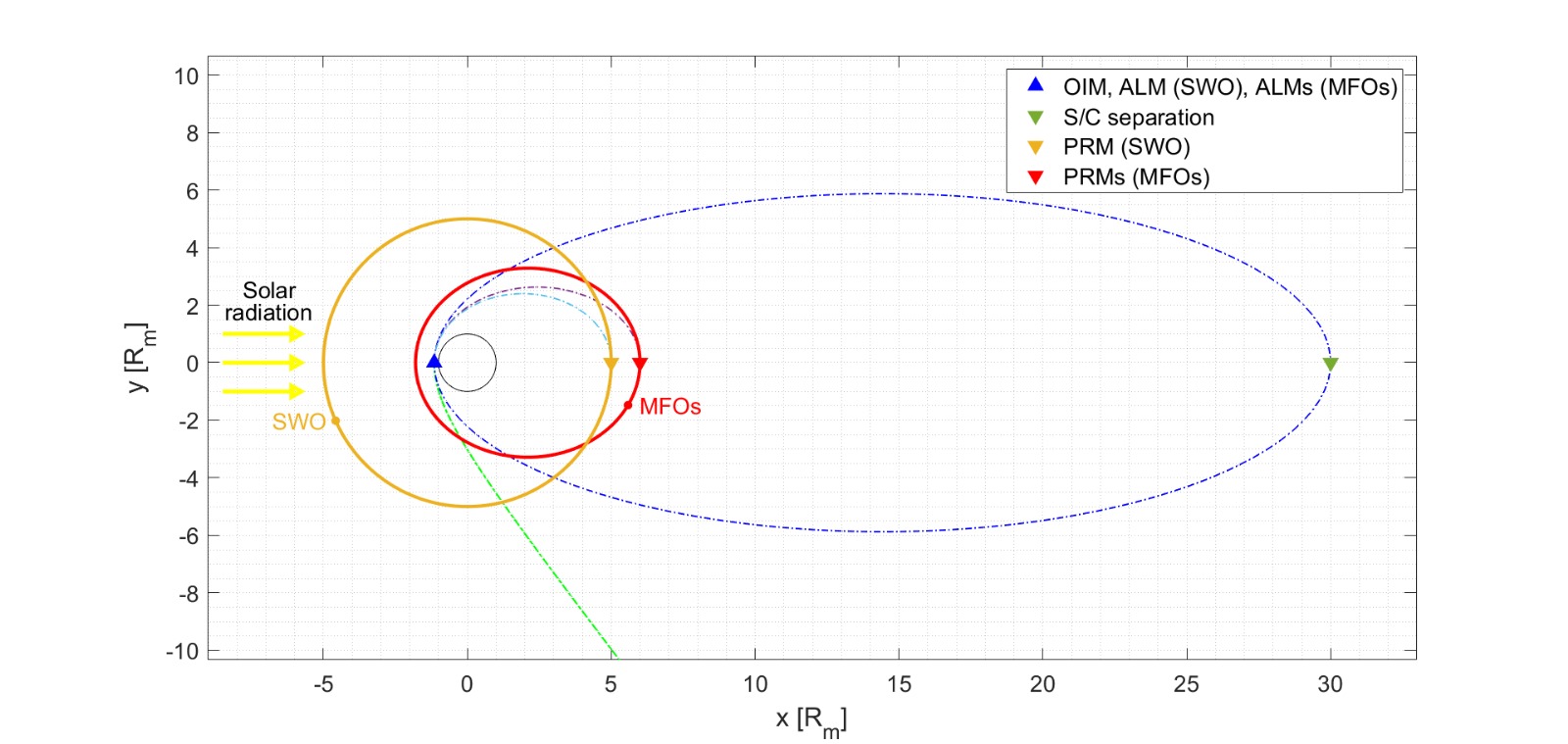}
\caption{\textcolor{red}{Mission trajectory close to Mars in Mars-Solar-orbital coordinates. The approach trajectory of the five spacecraft is shown in green. At the end of the approach trajectory an Orbit Insertion Maneuver (OIM) is performed to reach the capture orbit show in blue. Following the OIM the spacecraft separate. From the capture orbit, SWO lowers first its apoapsis, and finally increases its periapsis to reach its circular target orbit ($\mathrm{5 R_m \times 5 R_m}$) shown in orange. After the SWO has finished its maneuvers, the MFOs lower their apoapsis and raise their periapsis to reach their target orbit ($\mathrm{1.8 R_m \times 6 R_m}$) shown in red. The inclination of the orbital plane is $150^{\circ}$ for all orbits. A more detailed description of the maneuvers is provided in \cref{sec:orbit}.}
}
\label{fig:traj}
\end{figure*}

After launch, the five spacecraft will fly \textcolor{black}{in a stacked configuration along} a heliocentric elliptic transfer orbit to Mars. The approach trajectory along with the final orbits of the spacecraft and the transfer orbits needed to reach them are illustrated in \autoref{fig:traj}. Initially the four MFOs are stacked on top of the SWO. In this transit configuration the spacecraft will perform a \textcolor{red}{number of Trajectory Correction Maneuvers (TCMs)} before reaching Mars' sphere of influence, arriving at a periapsis of \textcolor{red}{$\mathrm{1.15 R_m}$} with an inclination of $150^{\circ}$. \textcolor{red}{In the stacked configuration, the spacecraft perform an Orbit Insertion Maneuver (OIM) that brings them to a capture orbit with a periapsis of $\mathrm{1.15 R_m}$ and an apoapsis of $\mathrm{30 R_m}$. The low periapsis and high apoapsis of the capture orbit is chosen to maintain the propellant mass of the SWO within feasible limits set by the size of the SWO inside the launcher fairing.}

\textcolor{red}{Approaching the apoapsis of the capture orbit, approximately \qty{48}{\hour} after the OIM, all five spacecraft separate mechanically from each other. The early separation of the spacecraft is, again, a trade-off between the limited SWO propellant mass and an increase in mission operations complexity that arises from individual maneuvering of the spacecraft. Once all spacecraft reach the periapsis following the separation, the SWO performs an Apoapsis Lowering Maneuver (ALM) to bring it to a $\mathrm{1.15 R_m \times 5 R_m}$ orbit. As soon as the SWO reaches the apoapsis of this new orbit, it will further perform a Periapsis Raise Maneuver (PRM) to circularize its orbit to its target orbit ($\mathrm{5 R_m \times 5 R_m}$). The MFOs, in contrast, continue an additional rotation along the capture orbit to avoid performing maneuvers simultaneously with the SWO. Once the MFOs reach the capture orbit periapsis again, they perform simultaneous ALMs to obtain a $\mathrm{1.15 R_m \times 6 R_m}$ orbit. When the MFOs reach the apoapsis of this orbit, they perform PRMs to obtain their target orbit of $\mathrm{1.8 R_m \times 6 R_m}$. Finally, the MFOs perform a Formation Configuration Maneuver (FCM) to reach the required cartwheel helix formation. The $\Delta V$ required to perform the required orbital maneuvers and the propellant mass burned during the thrusts are presented in chronological order in \autoref{tab:dv}.} 

\begin{figure*}[h]
\centering
\includegraphics[width=0.9\linewidth]{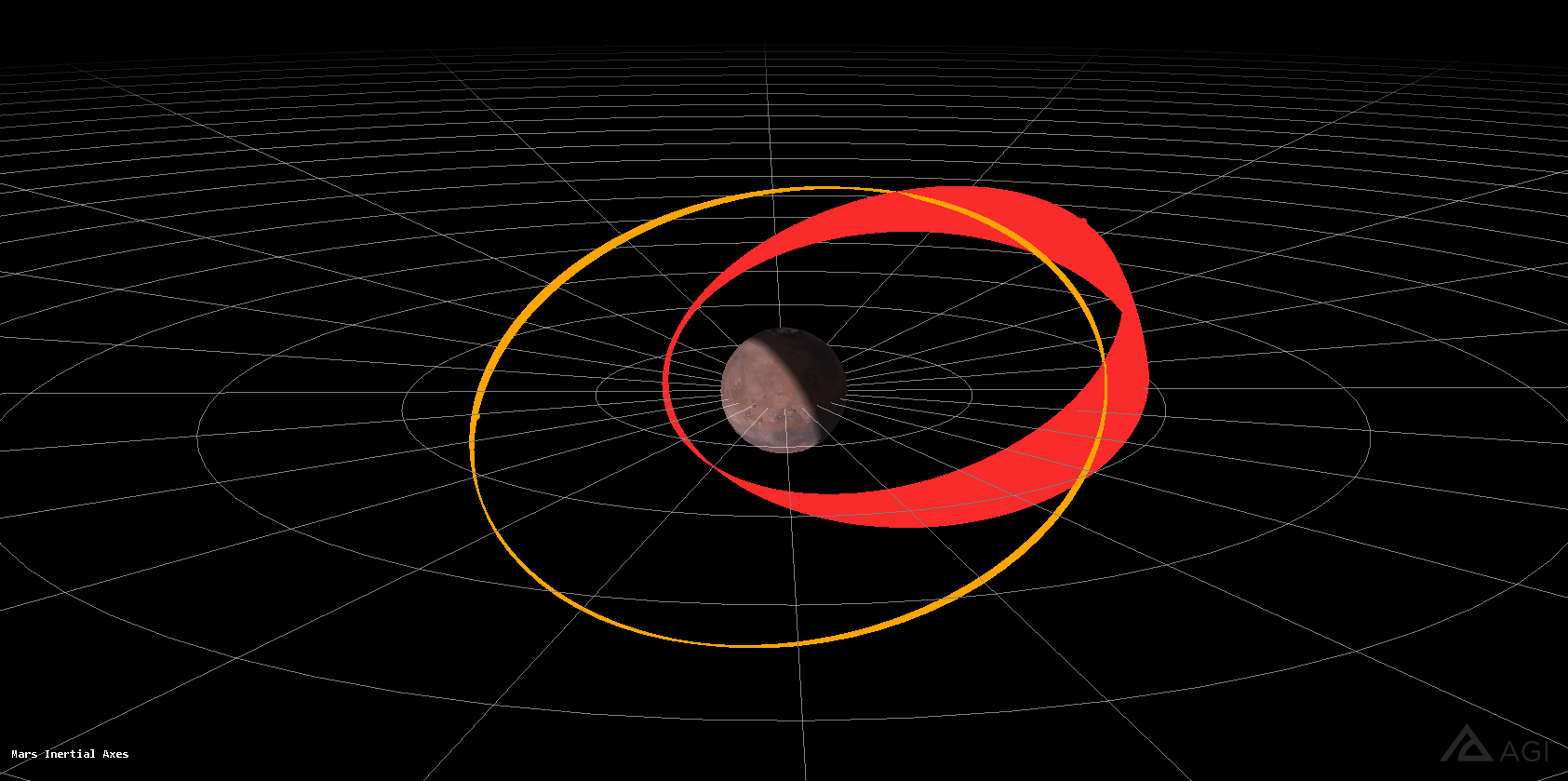}
\caption{Orbits propagated for 100 days. The orbit of the SWO is shown in orange, and the orbit of the MFOs in red. $\mathrm{J_2}$ perturbations will move the RAAN of the MFOs' orbit over time at a constant rate of $0.22^{\circ}$ per day. The figure is a screen capture from the STK simulation software.}
\label{fig:prop}
\end{figure*}

The choice of orbit for the MFOs ($\mathrm{1.8 R_m \times 6 R_m}$) satisfies the scientific requirement of orbiting in the magnetotail.  \textcolor{red}{The $150^{\circ}$ orbital inclination that all spacecraft maintain throughout the mission is chosen to maximize the benefit obtained from the $\mathrm{J_2}$ effect. Due to the optimized utilization} of the $\mathrm{J_2}$ effect, the time spent in the tail region is increased by a factor of five to $280$ days. A schematic of the orbit propagation can be seen in \autoref{fig:orbit}, and the simulated temporal evolution of the orbits is shown in \autoref{fig:prop}.

\begin{table*}[h]
\centering
\caption{$\Delta V$ budget. \textcolor{red}{The maneuvers are presented in chronological order. A symbol $\times$ indicates which spacecraft perform(s) the maneuver in question. Before spacecraft separation, the spacecraft are in a stacked configuration, and the SWO is responsible for the maneuvers. The spacecraft separation is performed mechanically and requires no propellant. The required $\Delta V$ and propellant mass is always indicated for a single spacecraft (or for the whole spacecraft stack prior to separation).}}  \label{tab:dv}
\textcolor{red}{
\begin{tabular}{ c c c c c c }
	\hline
Maneuver& $\Delta V$ [m/s]  & SWO & Each MFO & Propellant mass [Kg] \\
	\hline 
TCMs & 10.5 & $\times$ &   & 6.2  \\ 	
OIM & 808.2 & $\times$ &  & 467.7 \\
Spacecraft separation & -- & $\times$ & $\times$ & -- \\
ALM   &  392.3 & $\times$ &  & 65.7 \\
PRM   &  648.4 & $\times$ &  & 114.6 \\
ALM   &  321.51 &  & $\times$ & 22.7 \\
PRM   &  170.9 &  & $\times$ & 11.2 \\ 	
FCM   & 420   &  & $\times$ & 25.1     \\
    \hline
\end{tabular}
}
\end{table*}

\subsection{Orbit \& Attitude Maintenance}

In addition to propellant required for the $\Delta V$ to reach the required Martian orbits, propellant is budgeted for orbit maintenance and attitude control over the mission lifetime. Orbit Trim Maneuvers (OTMs) are required to maintain and fine tune the orbits. The propellant mass required for OTMs of each spacecraft is estimated based on the experience gained from the MAVEN mission \citep{jesick2017navigation}. Attitude Control Maneuvers (ACMs) augment the use of reaction wheels to adjust or maintain the attitude of the spacecraft. ACMs include periodical thruster firings for offloading torques from the reaction wheels to keep them out of saturation. The propellant allocated for OTMs and ACMs is \SI{21.1}{\kilogram} for the SWO and \SI{7.4}{\kilogram} for each MFO. Attitude control details and requirements are presented in \autoref{sec:attitude_requirements}.

\begin{figure*}[!b]
    \centering
    \includegraphics[width=0.9\linewidth]{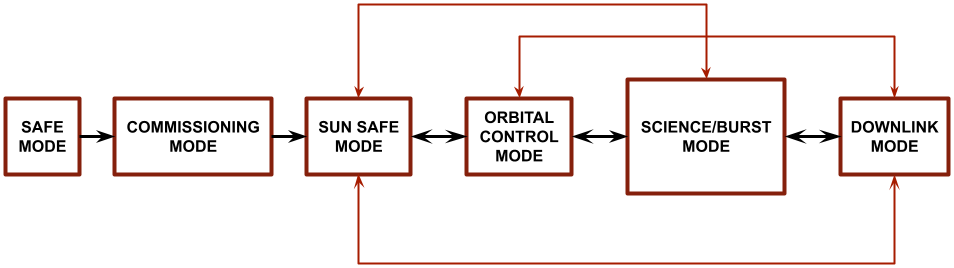}
    \caption{State Mode Diagram. Arrows depict the possible transitions between different modes. In general, any state mode is accessible directly from any other state mode. The exceptions are Safe Mode and Commissioning Mode, which are not used after they have been completed at the early phases of the mission. Sun Safe Mode acts as the contingency mode after launch.}
    \label{fig:states}
\end{figure*}

\subsection{Space Segment}
The space segment of the mission consists of the SWO and the four MFOs, which differ in design due to varying payloads and functionalities. The following subsections cover the space segment in more detail.

\subsubsection{Structure \& Spacecraft Design}
The primary structure of both types of spacecraft consists of a \SI{1.214}{\metre} cylindrical core that encloses the propellant tanks, made of titanium (Ti6AI4V STA). Exterior panels are attached to the central core. An aluminium honeycomb sandwich structure with graphite composite face sheets is used for all the primary structure elements \textcolor{black}{of both configurations}, providing enough stiffness to sustain the launch loads and induced vibrations. The panels sections are joined with bonded composite L-brackets. The general dimensions of the SWO spacecraft are \SI{2.3}{\metre} \texttimes\,\SI{2.3}{\metre} \texttimes\,\SI{1.8}{\metre}, \textcolor{black}{whereas the MFOs have a diameter of \SI{1.5}{\meter} and a height of \SI{1.2}{\metre}}. The preliminary \textcolor{red}{dry} mass of the structure \textcolor{red}{alone} is estimated to be \SI{240}{\kilogram} for the SWO \textcolor{black}{and \SI{90}{\kilogram} for each MFO}.
The material structure and structure layout employed is widely used in space missions \citep{YASAKA2003449}. This provides a high TRL, and heritage e.g. from the Dawn \citep{Dawn_Thomas2011} and MAVEN \citep{Jakosky2015} spacecraft for the SWO and Cluster \citep{Escoubet1997} for the MFOs.

\begin{figure*}[!ht]   
     \centering
     \includegraphics[width=0.95\textwidth]{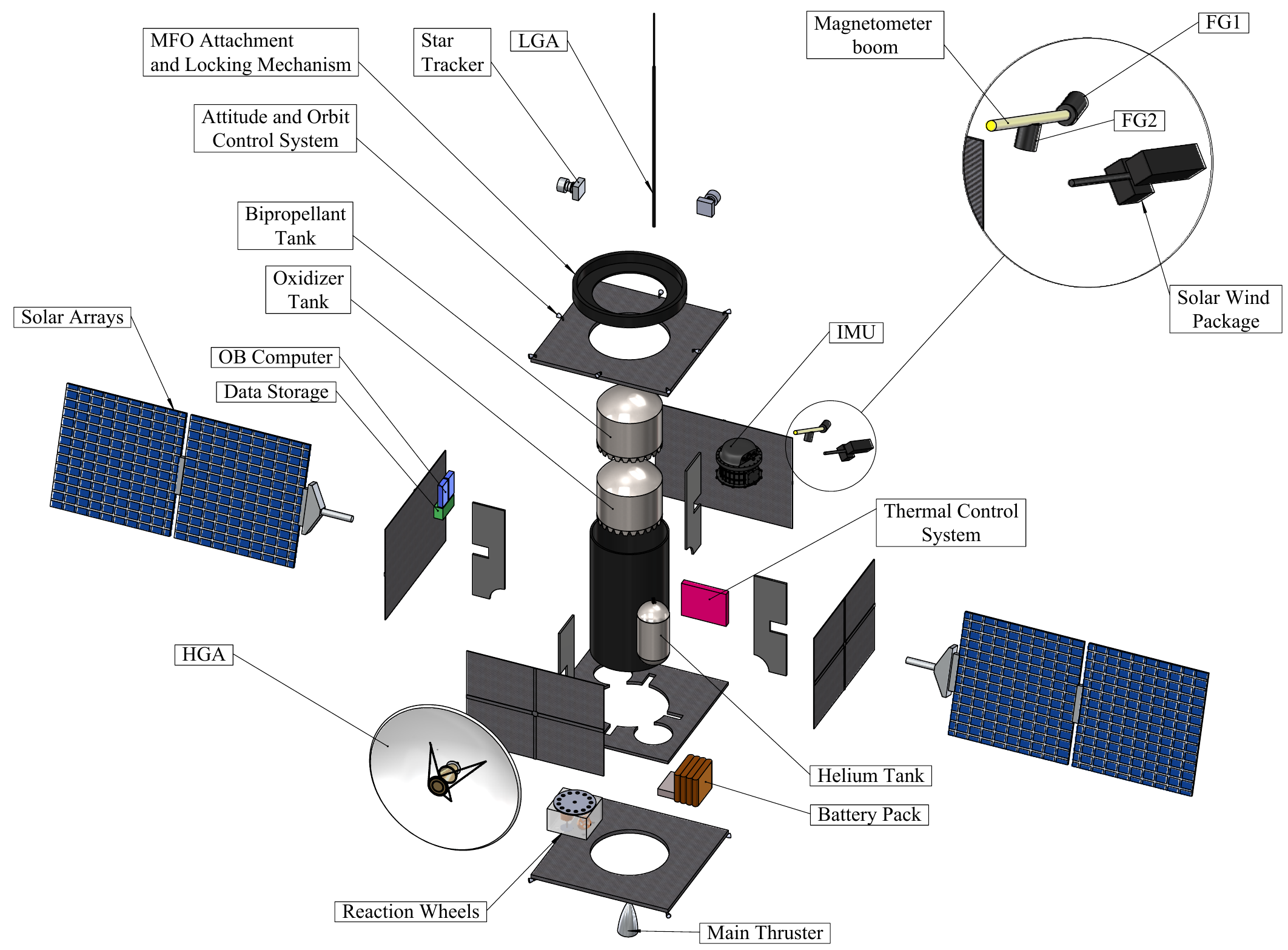}
     \caption{Expanded view of the Solar Wind Observatory and all major subsystems. \textcolor{red}{Some small-sized subsystems are scaled up for improved visualisation.}}
     \label{fig_structure}
\end{figure*}

In the bottom part of the spacecraft, a central cylinder is used to ensure precise attachment to the payload adaptor. On the top part of the spacecraft, an attachment and locking mechanism is used. The MFOs are stacked on top of each other using the aforementioned locking mechanism which will be designed in further mission design phases. An exploded view of the SWO with major subsystems is presented in \autoref{fig_structure}.

\subsubsection{Mass Budget}
To calculate the mission mass budget, the mass of each subsystem was derived based on estimates and data on existing subsystems. \textcolor{black}{The estimated payload masses are presented in \cref{tab:payload_resources}.} A margin of \SI{5}{\percent} to \SI{20}{\percent} was added to the calculated mass of each subsystem. Moreover, an additional overall system margin of \SI{20}{\percent} was added to the sum of subsystem masses to obtain the final dry mass estimate of the system. The total wet mass of the system was obtained by adding up the dry mass and the required propellant mass with margins. The margin philosophy is explained in \autoref{sec:margins}. The mass budget that shows the masses of each spacecraft and the total system mass is presented in \autoref{tab:massbudget}.

\begin{table}[H]
\centering
\caption{Final mass budget}  \label{tab:massbudget}
\begin{tabular}{ c c c c }
\hline
Spacecraft & SWO [kg] & 1 MFO [kg] & Margin \\
	\hline
Dry mass & \textcolor{red}{517} & \textcolor{red}{182} &  -- \\

Dry mass (marg.) & \textcolor{red}{621} & \textcolor{red}{218} & 1.20 \\

Propellant (marg.) & \textcolor{red}{730} & \textcolor{red}{69} & 1.10 \\

Total mass & \textcolor{red}{1364} & \textcolor{red}{288} & -- \\

\textcolor{red}{\SI{2516}{\kg}} & -- & -- & -- \\
\hline
\end{tabular}
\end{table}

\subsubsection{State Modes}
The SWO and MFO will operate in seven different main state modes presented in \autoref{fig:states}. The different state modes are designed for different phases of the mission. At the beginning of the mission, during launch and part of the transit, the system will stay in Safe Mode. This is a low power mode where as many subsystems as possible are turned off, and special safety measures are taken to ensure they will not turn on unexpectedly in any critical phase at the start of the mission. In addition, unintended separation of the spacecraft from each other should be strictly prevented.

From Safe Mode the system will proceed to Commissioning Mode, where e.g. solar panels are deployed in order to start power generation and health checks are performed on the instruments. Sun Safe Mode is entered after commissioning for the duration of the transit. It ensures that the system generates power, but payloads stay powered down or in a low power mode. Orbital Control Mode is entered as the spacecraft arrives at Mars. This mode enables orbital maneuvering utilizing the thrusters of the spacecraft. The mode is critical for reaching the desired orbits of the spacecraft, and performing small corrective maneuvers later on during the mission.

When the required orbits are reached, the spacecraft can proceed to start the science phase of the mission by operating in Science Mode. In this mode the spacecraft are designed to operate all of their instruments in order to collect data. 
At specific events during the mission, e.g. boundary crossings, the so-called Burst Mode can be initiated to enable short periods of increased data acquisition rates for the instruments. \textcolor{black}{Science operations are not allowed in Safe Mode or during data transmission.}

For transmitting the generated data, each spacecraft can enter Downlink Mode. For the MFOs this enables data transmission to the SWO. Furthermore, the SWO is able to downlink the self-generated data and the data received from the MFOs to the ground station on Earth. Receiving is activated in most state modes to enable commands to be sent to the spacecraft. The only exceptions are Safe Mode and Sun Safe Mode during transit, where only the SWO is receiving, as the spacecraft are still attached together.

In the following sections, Safe Mode and Sun Safe Mode can together be referred to as "safe modes", whereas "nominal modes" refer to all other operating modes.

\subsubsection{Power Budget} \label{sec:power_budget}

The power budget\textcolor{red}{s} of the spacecraft ha\textcolor{red}{ve} been designed by assuming \textcolor{red}{worst case solar irradiance conditions, as well as} end-of-life conditions for different parts of the power system. This means that e.g.~the degradation of solar cells and batteries over the mission lifetime has been accounted for when sizing the system. \textcolor{black}{The estimated power consumption of each payload can be found in \cref{tab:payload_resources}.} The total power consumption of the SWO in nominal state modes at the Red Planet will range from a maximum of \SI{440}{\watt} (Downlink Mode) to \SI{240}{\watt} (other nominal modes). The power generated by the SWO's solar panels in the Sun will be \SI{400}{\watt} at Mars. In contrast, the total power consumption of the MFO will vary between \SI{250}{\watt} (Downlink Mode) and \SI{150}{\watt} (other nominal modes). The power generated in the Sun by an MFO at Mars will be \SI{250}{\watt}.

All nominal state modes of a spacecraft, except Downlink Mode, consume the same amount of power. This results from sufficient heat dissipation being the restricting factor that determines the lower limit for power consumption. The reason for the higher power consumption of Downlink Mode is that in addition to the heat required to maintain the thermal balance of the satellite, some power is also radiated away from the satellite in transmission. Furthermore, for the SWO, Downlink Mode is considered in two separate submodes: transmitting to Earth, or transmitting to the MFOs. When transmitting to the MFOs, the SWO can use its payloads without compromising the thermal or power budget.

\begin{figure*}[h!] 
  \centering
  \begin{subfigure}[b]{0.48\textwidth}
    \includegraphics[width=8cm]{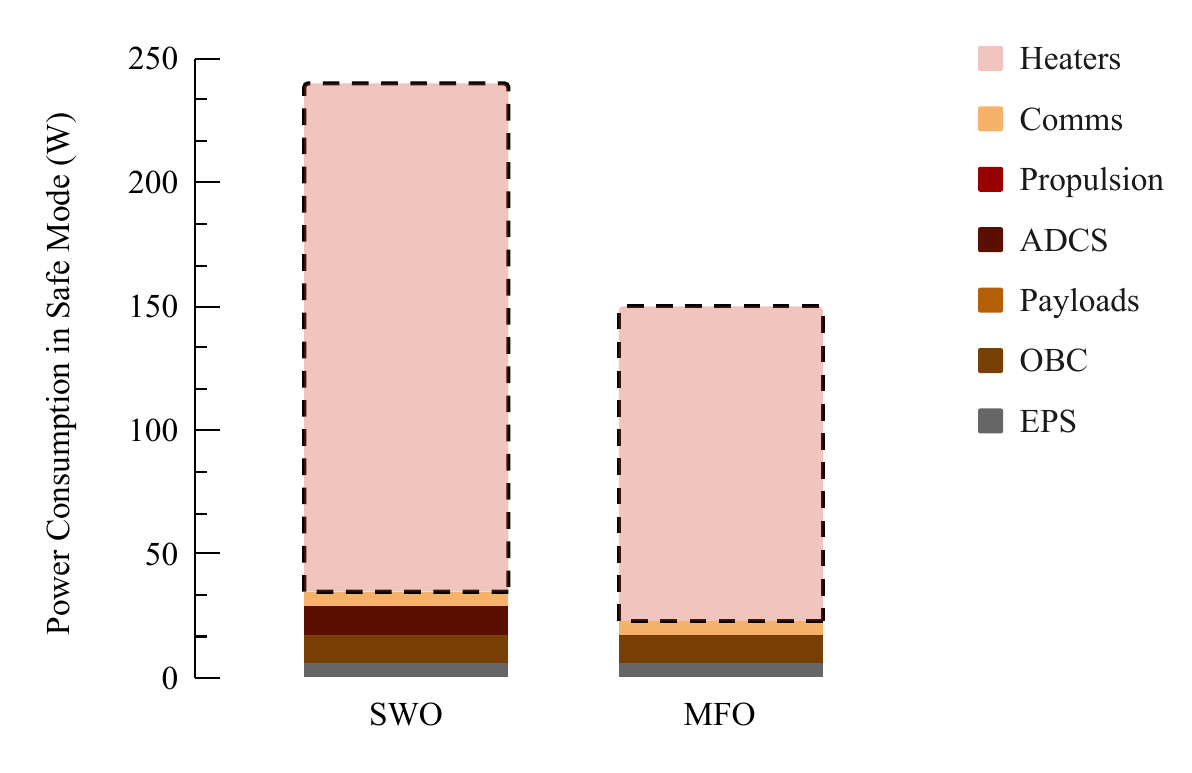}
    \caption{Safe Mode} \label{fig:power_ssafe}
  \end{subfigure}
  \hfill
  \begin{subfigure}[b]{0.48\textwidth}
    \includegraphics[width=8cm]{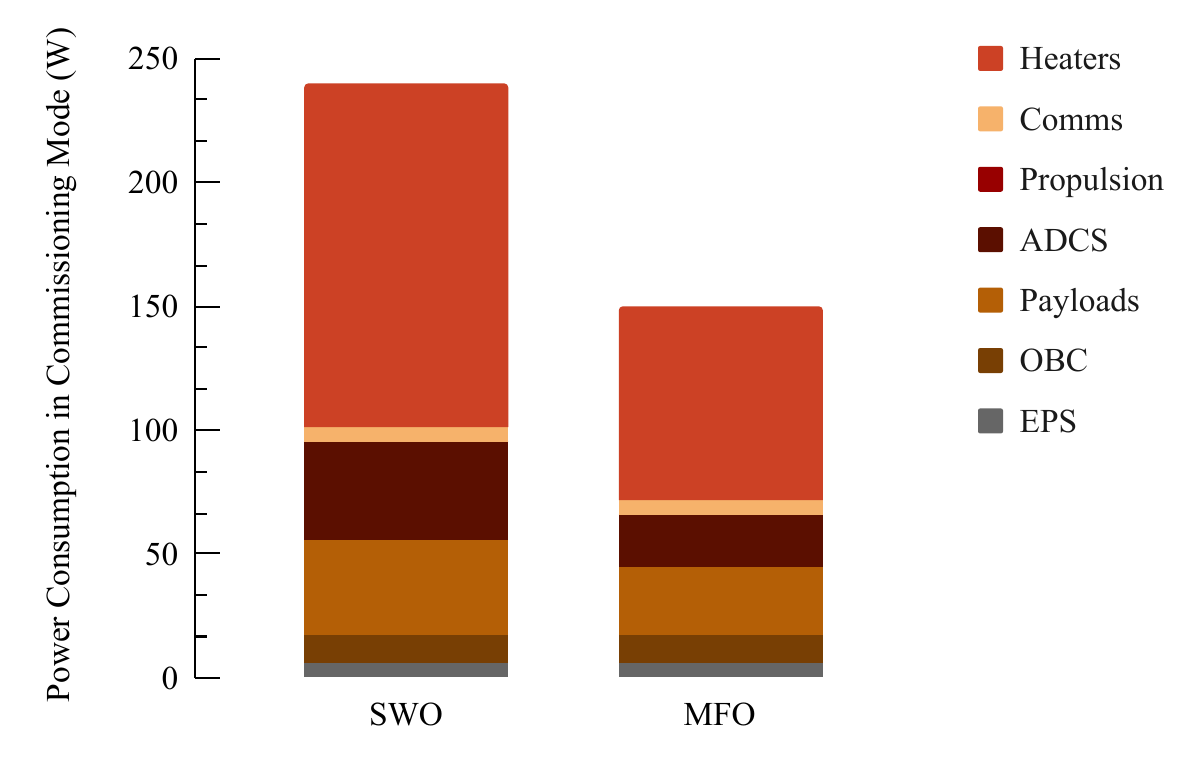}
    \caption{Commissioning Mode} \label{fig:power_commissioning}
  \end{subfigure}
  \hfill
    \begin{subfigure}[b]{0.48\textwidth}
    \includegraphics[width=8cm]{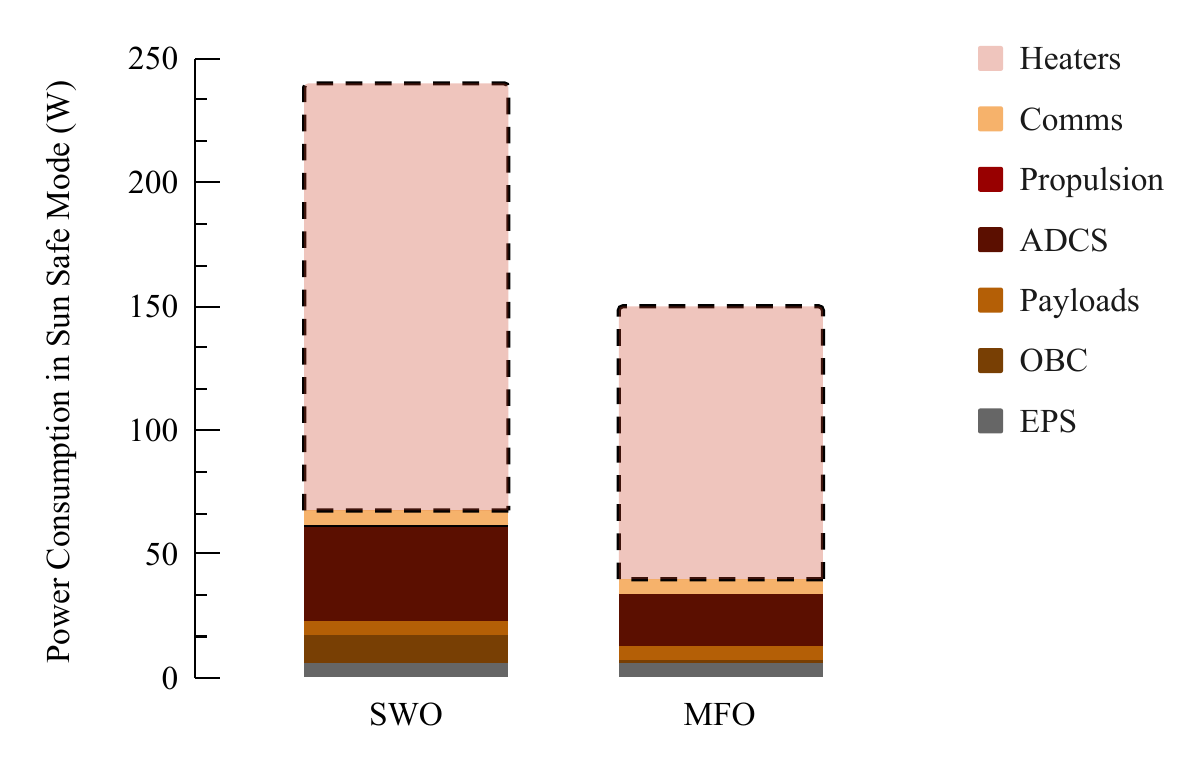}
    \caption{Sun Safe Mode} \label{fig:power_sunsafe} 
  \end{subfigure}
    \hfill
  \begin{subfigure}[b]{0.48\textwidth}
    \includegraphics[width=8cm]{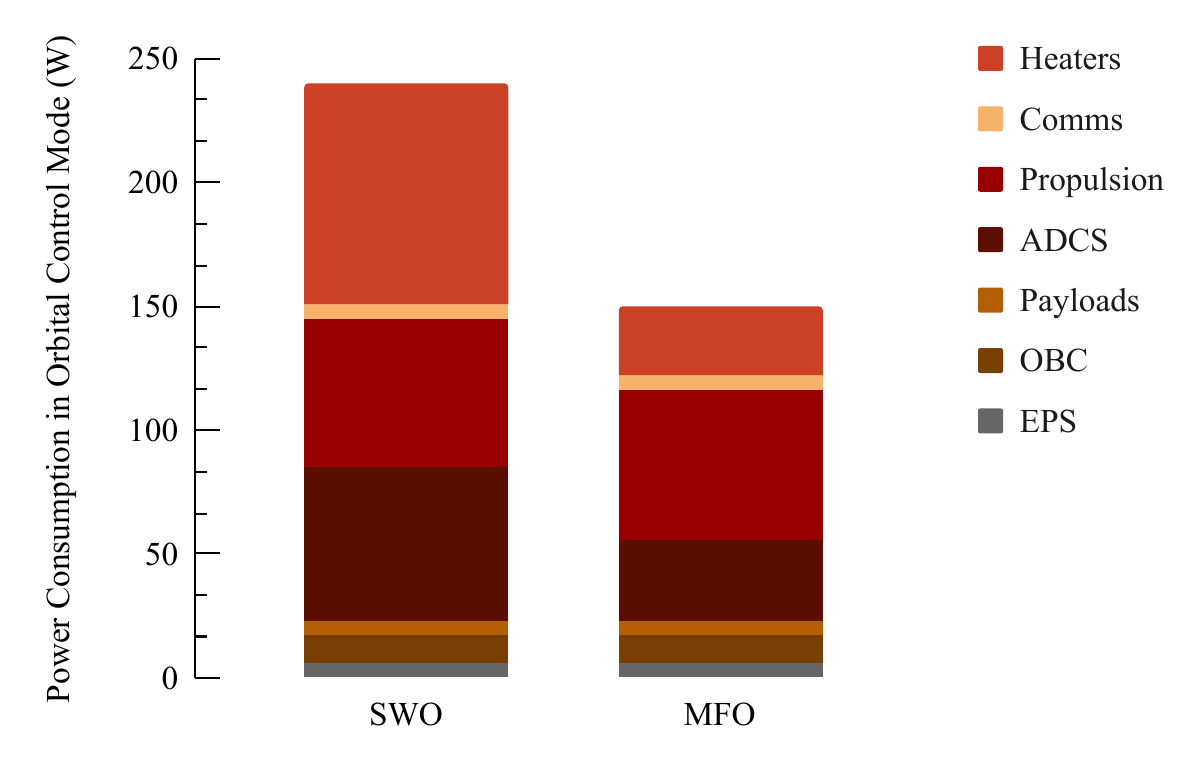}
    \caption{Orbital Control Mode} \label{fig:power_orbital}
  \end{subfigure}
  \hfill
    \begin{subfigure}[b]{0.48\textwidth}
    \includegraphics[width=8cm]{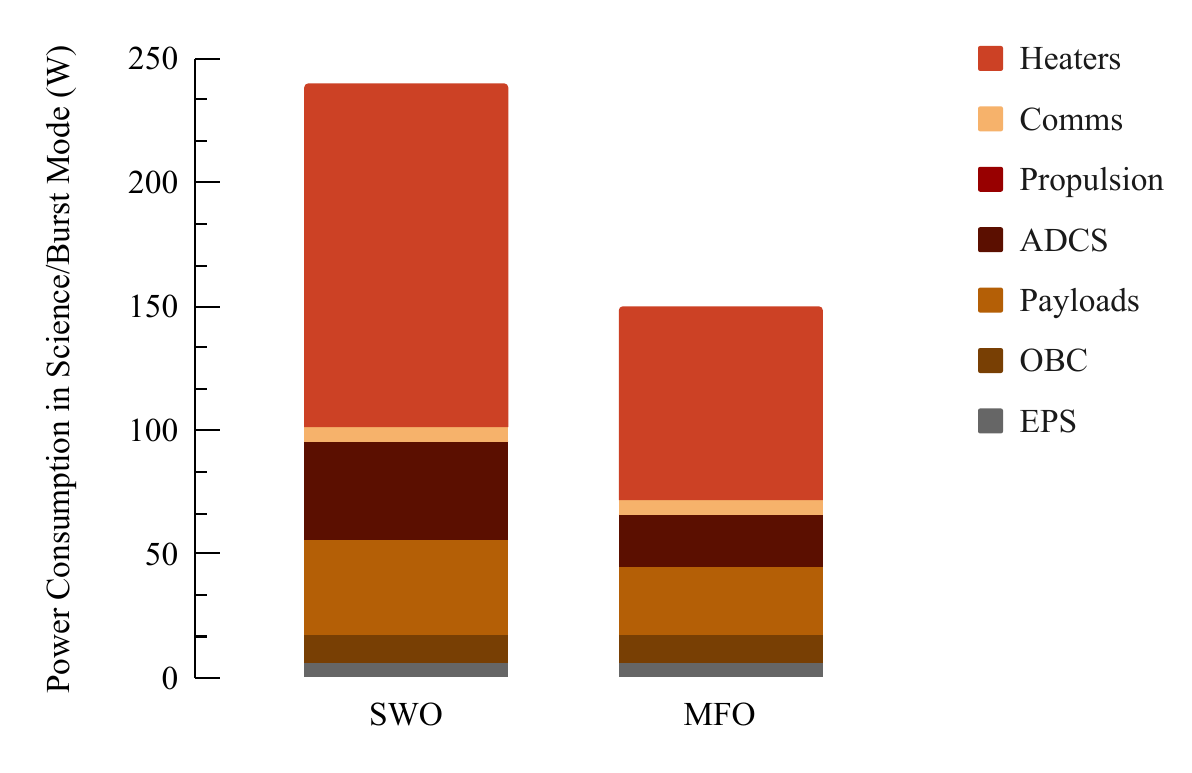}
    \caption{Science/Burst Mode} \label{fig:power_science}
  \end{subfigure}
    \hfill
  \begin{subfigure}[b]{0.48\textwidth}
    \includegraphics[width=8cm]{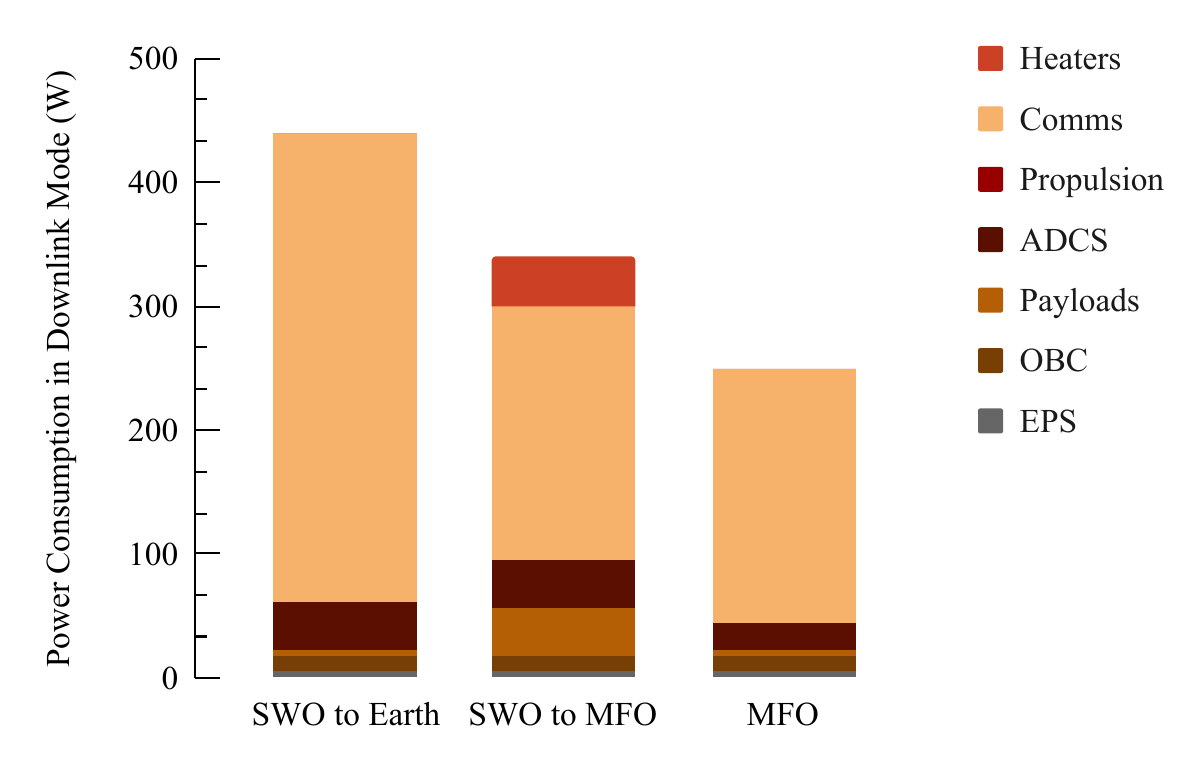}
    \caption{Downlink Mode} \label{fig:power_downlink}
  \end{subfigure}
  \caption{Power consumption in different state modes of the SWO and an MFO. Note the different scale of the vertical axis for Downlink Mode. In addition, note that in Safe Mode and Sun Safe Mode, the total power consumption may be lower than the total shown in the figure. The uncertain part is illustrated with a lighter box surrounded by a dashed line. The power budget is presented in detail in \autoref{sec:power_budget}}. \label{fig:power_budget}
\end{figure*}

In the safe modes, Safe Mode and Sun Safe Mode, the power consumption can potentially be lower than in nominal state modes. For example, during transit in Sun Safe Mode, the spacecraft are closer to the Sun than they are at Mars, and the required heating power produced by the spacecraft is lower. Additionally, if the power balance of a spacecraft would become compromised during nominal operations at Mars, the Sun Safe Mode can be initiated in order to save power while waiting for the batteries to recharge. The power consumption of different state modes is illustrated in \autoref{fig:power_budget}.

In the safe modes, the main factor limiting how low the power consumption can be decreased is the requirement to maintain the thermal balance of the spacecraft on a level that does not harm the spacecraft or their subsystems. The required power can be minimized, if the most temperature sensitive components are placed close to each other, and they are thermally well isolated from the environment. However, the tentative thermal modelling of the spacecraft does not enable detailed estimations of the power consumption in the safe modes during different mission phases. The detailed analysis of the power consumption in the safe modes will be performed in later mission design phases.

\textcolor{red}{The estimated maximum eclipse time during the mission is \qty{71}{\minute} for the SWO, and \qty{112}{\minute} for the MFOs.} The designed solar array power generation capacity is sufficient to charge the batteries of both types of spacecraft between the eclipses while staying in nominal operation modes. Without accounting for Downlink Mode, power is produced with a margin of approximately \SI{50}{\percent} compared to the other nominal state modes. Accounting for the higher power consumption of Downlink Mode reduces the margin significantly, but battery capacity is sized to enable the downlink sessions required during the mission (see \autoref{sec:telemetry}). The batteries used for the SWO and each MFO are \SI{3000}{Wh} and \SI{1500}{Wh} silver-cadmium batteries respectively. If, for any reason, the power balance of any of the spacecraft would become compromised, the Sun Safe Mode can be initiated in order to save power while waiting for the batteries to recharge.

\subsubsection{Thermal Budget}

For thermal modelling of the spacecraft, a coarse overall spacecraft thermal mathematical model (TMM) was utilized. The tentative modelling shows that to stay inside the estimated nominal operating temperature range with margins (\SI{-20}{\degreeCelsius} to \SI{60}{\degreeCelsius}), the SWO and each MFO require a continuous average heat dissipation of \SI{240}{\watt} and \SI{150}{\watt} respectively. As subsystem heat dissipation alone does not reach the required level, heaters are used to generate the required total heat. In addition, multi-layer insulation (MLI) is considered for thermal insulation of the spacecraft. No active cooling is required to maintain the spacecraft temperature according to this estimate, provided sufficient heat transfer within the spacecraft to even out internal thermal gradients. At later system design phases, a more sophisticated thermal control scheme could be devised to optimize the power consumption and thermal stability of the spacecraft. As of now, the feasibility of the thermal budget has been demonstrated by assuming simple constant thermal dissipation power.

As all power produced by the subsystems on-board the spacecraft (except power radiated from the antennas in Downlink Mode) is assumed to be dissipated as heat in the spacecraft, the total heat dissipation budgets are equal to the power budgets in each operating mode (except Downlink Mode). In Downlink Mode, the heat dissipation of the SWO is \SI{200}{\watt} lower than the power consumption. Similarly, the heat dissipation of a MFO is \SI{100}{\watt} lower than its power consumption in Downlink Mode.

\subsubsection{Telemetry Budget \& Telecommand} \label{sec:telemetry}
In addition to performing scientific measurements, the SWO serves as a communication relay between the MFO formation and the ground segment on Earth. For this purpose, the SWO carries a high gain dish antenna (HGA) with a diameter of \SI{2.5}{\metre}. The X-band is chosen for the data link between Earth and Mars, similarly as has been done for instance on the Mars Reconnaissance Orbiter \citep{GRAF2005566}. The strict pointing requirement of the HGA (\textless\,\ang{0.3}) is achieved by pointing the antenna semi-independently from the spacecraft body. To enable communications between the SWO and the MFOs, each of the five spacecraft carries a low gain dipole antenna (LGA) that poses no strict pointing requirements. Communication between the MFOs and the SWO will use the S-band frequency range, which was shown by link calculations to be suitable for the intersatellite link.

The link budget of the mission is heavily dependent on the mutual distances between the spacecraft, as well as the distance of the SWO from Earth. The simulated best and worst case distances, as well as the average distance over time, are presented in \autoref{tab:distances}. The corresponding link budgets are detailed in \autoref{tab:linkbudget}. The significant variance in downlink rates is attributed to differences in free-space path loss (FSPL) that depends on the distance between the transmitter and receiver. FSPL grows rapidly as distance $d$ between the transmitter and the receiver increases (FSPL $\propto d^2$), and leads to signal attenuation.

\begin{table}[h]
\centering
\caption{Mutual distances during the mission. \textcolor{black}{The mean distances are weighted by time.}}  \label{tab:distances}
\begin{tabular}{ c c c c }
	\hline
& Min. & Max. & Mean \\
	\hline
SWO/Earth & \SI{5.7e7}{\kilo\metre} & \SI{3.2e8}{\kilo\metre} & \SI{1.5e8}{\kilo\metre} \\ 	
MFO/SWO& \SI{1.2e3}{\kilo\metre} & \SI{3.7e4}{\kilo\metre} & \SI{2.0e4}{\kilo\metre}  \\ 	
    \hline
\end{tabular}
\end{table}

\begin{table}[h]
\centering
\caption{Link budget as achievable downlink/uplink data rates \textcolor{black}{that correspond to the distances specified in \autoref{tab:distances}}.}  \label{tab:linkbudget}
\begin{tabular}{ c c c c }
	\hline
Direction & Min. & Max. & Mean \\
	\hline
SWO $\rightarrow$ Earth  & \SI{0.72}{Mbps} & \SI{24}{Mbps} & \SI{3.5}{Mbps} \\ 	
Earth $\rightarrow$ SWO  & \SI{2.1}{Mbps} & \SI{67}{Mbps} & \SI{9.9}{Mbps} \\ 	
MFO $\rightarrow$ SWO & \SI{6.4}{kbps} & \SI{6.2}{Mbps} & \SI{22}{kbps} \\ 	
SWO $\rightarrow$ MFO & \SI{6.4}{kbps} & \SI{6.2}{Mbps} & \SI{22}{kbps}  \\
    \hline
\end{tabular}
\end{table}

A majority of the proposed scientific heritage instruments (see \autoref{sec:payload}) enforce lossless compression on their measurement data or stream continuously low resolution data while storing high resolution data to be transmitted only on demand. The maximum estimated total data volume produced by the instruments is presented in \autoref{tab:datarate}. \textcolor{black}{The result is based on the estimated data rates of each payload detailed in \cref{tab:payload_resources}.} The data rate estimations are designed to account for both nominal Science Mode operations and higher data rate Burst Mode measurements. A significant margin of \SI{50}{\percent} has been added to the tentative estimations that are based on data rates specified for the proposed heritage instruments.

\autoref{tab:downlink_times} shows estimated downlink times for the amount of data produced during an average \SI{24}{\hour} period of mission operations. The downlink times are estimated between the different spacecraft, as well as between the SWO and the ground station network. The SWO achieves downlink times of \SI{3.4}{\hour} even in the worst case scenario, corresponding to a total of \SI{15}{\percent} of operation time on average. This enables downlinking all data produced by the SWO and the MFOs to Earth with good margin during the whole mission duration, independent from the mutual distance of Earth and Mars.

The MFOs, in contrast, require optimized downlink schedules to be able to transmit all science data to the SWO, as the worst case and mean downlink rates are too slow for efficient data transfer, but the best case downlink rate is excellent. The downlink sessions should be scheduled to take place when the distance between the MFOs and the SWO is close to minimum to ensure the downlink time is minimized. As the orbital periods of the SWO and the MFOs are \SI{18.6}{\hour} and \SI{12.8}{\hour} respectively, the spacecraft will undergo a sufficiently close encounter roughly every \SI{38}{\hour}. The amount of on-board data storage is sufficient to store the data produced over significantly longer periods of time than the time between adjacent downlink time slots (see section \autoref{sec:storage}). Thus, not all downlink opportunities have to be utilized. Downlink opportunities can occasionally be skipped, e.g. if the opportunities happen to occur during particularly interesting measurement possibilities, such as magnetotail border crossings or exceptional solar wind conditions.

The uplink times from the SWO to the MFOs or from Earth to the SWO will be short, since the transmitted data volumes are minor, as only short commands need to be transmitted in these directions. In addition, the uplink data rate from Earth is relatively high during the whole mission lifetime.

\begin{table}[h]
\centering
\caption{Maximum combined instrument data rate averaged over an orbit.}  \label{tab:datarate}
\begin{tabular}{ c c c c }
	\hline
Unit & Max. data rate & Duty cycle & Mean data rate \\
	\hline
SWO  & \SI{19}{kbps} & \SI{50}{\percent} & \SI{9.4}{kbps} \\ 	
MFO  & \SI{23}{kbps} & \SI{65}{\percent} & \SI{15}{kbps}  \\ 	
Total & \SI{112}{kbps} & -- & \SI{70}{kbps}  \\
	\hline
\end{tabular}
\end{table}

\begin{table}[h]
\centering
\caption{Downlink times \textcolor{black}{for the amount of data produced} over an average \SI{24}{\hour} period.}  \label{tab:downlink_times}
\begin{tabular}{ c c c c }
	\hline
Direction & Min. & Max. & Mean \\
	\hline
SWO $\rightarrow$ Earth & \SI{6}{\min} & \SI{3.4}{\hour} & \SI{42}{\min} \\ 
MFO $\rightarrow$ SWO & \SI{4}{\min} & \SI{57}{\hour} & \SI{17}{\hour} \\ 	
    \hline
\end{tabular}
\end{table}

\subsubsection{On-Board Computer and Data Storage} \label{sec:storage}
The radiation hardened RAD-750 onboard computer (OBC) proposed for the mission has heritage from several missions such as the Mars Reconnaissance Orbiter \citep{GRAF2005566} as well as the Curiosity \citep{6575245} and Perseverance \citep{9438337} rovers. As the SWO poses a major single point failure risk for the mission, the spacecraft is equipped with two redundant OBCs. The four MFOs are each equipped with a single RAD-750 OBC.

The onboard data storage \textcolor{red}{allocated for} each MFO is \textcolor{red}{\SI{30}{GB}}, whereas the SWO will carry \textcolor{red}{\SI{160}{GB}} of memory. The combined total data storage is designed to be sufficient for storing the total data produced by all spacecraft over an average \textcolor{red}{12 month} period. This is possible, as an MFO can store the data produced by itself over \textcolor{red}{6 months}, whereas the SWO can store the data produced by each MFO over \textcolor{red}{6 months}, as well as the data produced by itself over \textcolor{red}{12 months}. The amount of data storage contains \textcolor{red}{substantial} margin to enable \textcolor{red}{significant} \textcolor{black}{flexibility in} downlink scheduling (see \autoref{sec:telemetry}).

\subsubsection{Attitude Determination \& Control} \label{sec:attitude_requirements}
For attitude determination, each spacecraft will use \textcolor{red}{an inertial measurement unit (IMU) in combination with two star trackers. The star trackers are utilized for periodical IMU calibration, and they offer a redundant means of attitude determination}. The SWO carries four reaction wheels for standard attitude and pointing control and a total of twelve thrusters: one main thruster for orbital insertions and major orbital maneuvers accompanied by eleven smaller thrusters for attitude control and minor orbital maneuvers. Each of the spin stabilized MFOs will also carry twelve thrusters in a similar configuration.

The high gain antenna of the SWO requires a pointing to Earth with \textless\,\ang{0.3} error for downlink mode. The HGA can be pointed semi-independently from the rest of the SWO spacecraft body. The low gain dipole antennas of all the spacecraft are required to maintain an alignment with the normal of the orbital plane with \textless\,\ang{30} of error in order to obtain a data link between the SWO and the MFOs.

During science mode operations, the solar wind observing instruments of the SWO require a pointing accuracy of \textless\,\ang{10} towards the incoming solar wind. The MFOs are required to spin in orbit in order to extend their wire booms. The measurements do not impose any pointing requirements on the MFOs.

\subsubsection{Electromagnetic Interference Considerations} \label{sec:EMI}
As accurate and high resolution measurements of the Martian magnetosphere are key to the scientific goals of the mission, strict magnetic cleanliness of the spacecraft will be necessary to prevent unwanted interference from impacting measurements. \textcolor{red}{A key measure taken to reduce the magnetic disturbances caused by the spacecraft is to ``back wire" the solar panels. The back wiring method reduces solar panel current loops, and consequently the magnetic field disturbances induced by the loops. The method has successful heritage from missions such as Mars Global Surveyor \citep{acuna1996magnetic} and MAVEN \citep{Jakosky2015}.}

To limit the influence of \textcolor{red}{remaining} spacecraft-induced magnetic fields on the measurements, all fluxgate magnetometers are placed on \textcolor{red}{\SI{4.5}{\m}} long booms. Additionally, each spacecraft has two magnetometers on the same boom to allow for \textcolor{black}{cleaning of magnetic field data}. The primary scientific magnetometer is placed on the tip of the boom, whereas the second one, closer to the spacecraft body, acts as an auxiliary magnetometer that assists in identifying and removing potential magnetic interference by the spacecraft from the data. This approach has previously been employed e.g. on the Cluster mission \citep{balogh1997cluster}.

Electromagnetic interference  must be considered also from a communications perspective to ensure the spacecraft are not producing interference on their communication frequencies in the S- and X-bands.

\subsection{End-of-life \& Planetary Protection}
ESA missions are required to abide by planetary protection standards. M$^5$ would be classed as a Category III mission by the relevant planetary protection standard \citep{ecss_planet}. Therefore, this mission will inventorise and retain samples of organic materials used in the spacecraft, comply with bioburden requirements, and assemble the spacecraft in a cleanroom of ISO class 8 or above. \textcolor{black}{The mission is also required to have an impact probability $\leq 1\times10^{-4}$ for 50 years after launch to comply with the COSPAR planetary protection policy \citep{kminek2015cospar}. We compare our orbit parameters with \citet{SUCHANTKE2020} and conclude that there is a negligible probability of de-orbiting within 50 years.}

\section{Programmatics}\label{sec:programmatics}

\subsection{\textcolor{red}{Cost Estimate, Descoping Options and Additional Instrumentation}} \label{section:costs}

We expect M$^5$ to be classified as an L-class mission according to the Cosmic Vision strategy of ESA. We have not made detailed cost estimates, but we expect that meeting the cost limit of MEUR 1000 will be challenging. One area for cost reduction, which is not required but may be desirable, is the possibility of collaborating with international partners.

Given the significant cost of the mission, descoping options are possible at the cost of reducing the scientific objectives. From the MFOs, one or more spacecraft could be descoped to lower mass and cost. However, this would significantly hinder the fulfillment of the science objectives, as a 4 spacecraft formation is needed to achieve most science objectives, namely O1.1, O1.2, O1.3, O2.2 (see \autoref{tab_instrumentation}). A reduction to 3 spacecraft would reduce the 3D picture to a 2D picture, meaning that boundary orientation and movement could no longer be separated. In addition, the curlometer and wave telescope techniques would only give good scientific return in a limited number of cases. A further reduction to 2 spacecraft would make answering of the science questions even more challenging, reducing the data to a 1D picture. 

\textcolor{red}{In the initial, preliminary design presented in this study, all MFOs are designed the same. This reduces cost and adds instrument/measurement redundancy  for some instruments. It also provides additional possibilities of scientific observations and adds to spatial resolution and thus increases the scientific value of the overall mission. However, as given by the traceability of the instrument requirements in \autoref{tab_instrumentation}, there are possibilities to descope instruments onboard the MFOs without loss of science objectives presented in \autoref{tab_scienceobjs}, such as two of the electron spectrometers. Additionally, the absence of electric antennas on the MFOs would result in a limited loss of scientific objectives. Instead of descoping, replacement by other instruments could be considered. Some examples of instrumentation that would increase the scientific value of the mission are for example a radiation monitor such as the BepiColombo Environment Radiation Monitor (BERM) \citep{Pinto2022} or a solar energetic particle detector such as that in the Solar Intensity X-Ray and Particle Spectrometer (SIXS) onboard BepiColombo \citep{Huovelin2020}. This would for example assist in monitoring solar eruptive events such as CMEs, which can strongly influence the Martian magnetosphere. Another, but possibly more demanding option in terms of resource allocation, is an Energetic Neutral Atom (ENA) imager. Although ASPERA-3 \citep{Aspera3} onboard Mars Express and MINPA \citep{konglingao-F} onboard Tianwen-1 are probing the ENA environment of Mars, open questions still remain \citep{Ramstad_2022}. Thus, an ENA imager would improve the understanding of the dynamics of the Martian plasma environment. The addition of any of these instruments without descoping other instruments would however greatly alter the complete mission design and increase cost significantly, as the current system budgets (especially telemetry and propellant) are already at their respective limits. Thus, such additions are not considered in more detail here. 
}

\subsection{Mission Readiness \& Risk Analysis}

All mission components have Technology Readiness Level (TRL) $\geq$ 6, so there are no significant technological risks to the mission. Some significant operational risks have been identified for the mission. One risk would be if either the communication with the SWO or with one (or more) of the MFO would be lost (resulting in the loss of some science objectives). In the case of losing the SWO, it may be possible to use MRO\textcolor{red}{, MAVEN, or the ExoMars orbiter} as a relay instead \citep{edwards2014replenishing}. Another risk would be a failed launch, as well as an error in the orbit insertion, both of which could result in a total loss of the mission. An error in the alignment of the MFO tetrahedron is also be a possible risk. The solar panels or the electric antennas not deploying would cause major difficulties for the mission. \textcolor{black}{Using the risk analysis methods outlined in \cite{ECSSrisk} we believe all of these risks can be classed as either low (1 in 1000 projects) or very low (1 in 10000 projects) risks, and are thus deemed acceptable.}

\subsection{Outreach}

Outreach is a key aspect for scientific space missions. As a scientific community there is a responsibility to inform taxpayers about how their money is being spent on research. Furthermore, outreach is a key driver for inspiring and encouraging young people to consider careers in \textcolor{black}{Science, Technology, Engineering and Mathematics (STEM)}. M$^5$ \textcolor{black}{will therefore have an associated} outreach program, \textcolor{black}{designed in accordance with current best practices. This would consist of a pre-launch program of online and in-person events to build excitement, and continue with press releases announcing key science results, and accompanying educational materials for schools, following the model of previous ESA missions \citep{heck_2003_astronomy,larslindbergchristensen_2007_the}.}

\vspace{-1mm}
\section{Conclusion}\label{sec:conclusion}
Through detailed preliminary analysis, we show the feasibility of a multi-spacecraft mission to Mars, aiming to extend and complement our understanding of the Martian induced magnetosphere. This understanding will further extend our comprehension of induced magnetospheric systems generally, and of their interaction with the solar wind. Atmospheres are important for the presence of life, and the escape of the Martian one will be better understood by the quantitative characterization of the magnetotail and of the processes taking place there. \\
In order to study these regions and phenomena on different scales, and in order to separate spatial and temporal variations without having to use imperfect a priori information, a three-dimensional picture of the bow shock, magnetic pile-up boundary as well as the magnetotail are achieved thanks to a four spacecraft configuration. The remaining spacecraft will complement the fleet of solar wind observatories in our solar system, crucial in order to provide better data for space weather applications. \\
We show the feasibility of these objectives through detailed analyses of the orbital dynamics, formation requirements, and budget constraints such as mass, power and communication. We give an overview of spacecraft design incorporating all critical systems, and show the availability of heritage instruments sufficient to achieve the desired science objectives. \\
The presented ambitious but feasible mission concept shows that a comprehensive study of the Martian magnetospheric system is possible, which is imperative for future human exploration of Mars. We show that M$^5$ would greatly advance our understanding of atmospheric escape, and give a crucial reference point for comparative studies of other solar system and exoplanetary induced magnetospheres.

\section*{Acknowledgments}

The authors acknowledge funding from the European Space Agency (ESA) and the Austrian Research Promotion Agency (FFG), which supported Summer School Alpbach 2022, on the theme ``Comparative Plasma Physics in the Universe''. 

We also acknowledge valuable discussions with the tutors of the Summer School Alpbach 2022, Carlos Pintassilgo and Brian Reville.

MB gratefully acknowledges support from the Norwegian Space Agency to attend the Summer School Alpbach 2022.

MB-S and DT gratefully acknowledge the support from the \"{O}sterreichische Forschungsf\"{o}rderungs GmbH (FFG) for participating in the Summer School Alpbach 2022.

PD gratefully acknowledges funding support from the Centre National d'\'Etudes Spatiales (CNES) for participating in the Summer School Alpbach 2022.

ADI gratefully acknowledges support from Agenzia Spaziale Italiana to attend the Summer School Alpbach 2022.

JG and LS gratefully acknowledge funding support from Deutsches Zentrum f\"{u}r Luft- und Raumfahrt for participating in the Summer School Alpbach 2022.

CJKL gratefully acknowledges support from SRON to attend the Summer School Alpbach 2022, and gratefully acknowledges support from the International Max Planck Research School for Astronomy and Cosmic Physics at the University of Heidelberg in the form of an IMPRS PhD fellowship. CJKL gratefully acknowledges further support by the German Deut\-sche For\-schungs\-ge\-mein\-schaft, DFG\/ in the form of an Emmy Noether Research Group -- Project-ID 445674056 (SA4064/1-1, PI Sander) and the Federal Ministry of Education and Research (BMBF) and the Baden-W\"{u}rttemberg Ministry of Science as part of the Excellence Strategy of the German Federal and State Governments. 

SL and SNÖ gratefully acknowledge support from the Swedish National Space Agency to attend the Summer School Alpbach.

VL gratefully acknowledges the support from the Finnish Centre of Excellence in Research of Sustainable Space (FORESAIL) and the School of Electrical Engineering of Aalto University for participating in the Summer School Alpbach 2022.





\bibliographystyle{jasr-model5-names}
\biboptions{authoryear}
\bibliography{refs}
\end{document}